\documentclass[a4paper, 11pt]{article}
\usepackage{amsthm}
\usepackage{multicol}
\usepackage[affil-it]{authblk} 
\usepackage{etoolbox}
\usepackage{lmodern}
\usepackage[a4paper]{geometry}
\usepackage[in]{fullpage}
\usepackage[utf8]{inputenc}
\usepackage{amsmath}
\usepackage{amsthm}
\usepackage{enumitem}
\usepackage{tcolorbox}
\usepackage{physics}
\usepackage{amssymb}
\usepackage{tablefootnote}
\usepackage{float}

\numberwithin{equation}{section}
\theoremstyle{definition}

\theoremstyle{remark}

\usepackage{booktabs}
\usepackage{amssymb}
\usepackage{cite}

\usepackage{amsfonts}
\usepackage{graphicx}
\usepackage[export]{adjustbox}
\usepackage{wrapfig}
\usepackage{sidecap}
\usepackage{ulem}
\usepackage[dvipsnames]{xcolor}
\usepackage[colorlinks]{hyperref}
\hypersetup{linkcolor=blue,citecolor=blue,urlcolor=blue}

\title{Metastrings, Metaparticles and Black Hole Thermodynamics: On the Road Towards a Non-singular Black Hole Remnant}

\makeatletter
\patchcmd{\@maketitle}{\LARGE \@title}{\fontsize{16}{19.2}\selectfont\@title}{}{}
\makeatother

\author[1,2]{Paul-Robert Chouha\thanks{Email: \href{mailto:paul.chouha@mcgill.ca}{paul.chouha@mcgill.ca}}}
\affil[1]{Department of Physics, Ernest Rutherford Physics Building, McGill University,\\
        3600 rue Universit\'e, Montr\'eal, Qu\'ebec H3A 2T8, Canada}
        
 \affil[2]{McGill University School of Continuing Studies,\\
680 Sherbrooke St. West, Montreal, Quebec Canada H3A 3R1}

\date{\vspace{-5ex}}

\newcommand{\tx}{\tilde{x}}

\newcommand{\tl}{\tilde{l}}
\newcommand{\tA}{\tilde{A}}
\newcommand{\tM}{\tilde{M}}
\newcommand{\tm}{\tilde{m}}
\newcommand{\tE}{\tilde{E}}
\newcommand{\tT}{\tilde{T}}
\newcommand{\ttau}{\tilde{\tau}}
\newcommand{\ttt}{\tilde{t}}

\newcommand{\tS}{\tilde{S}}
\newcommand{\tC}{\tilde{C}}
\newcommand{\tmu}{\mu}

\begin{document}
\maketitle

\begin{abstract}
We investigate the thermodynamic evolution and endpoint of black hole evaporation in the framework of metastring theory \cite{Freidel:2015pka, Freidel:2017xsi} and its particle excitations, the metaparticles \cite{Freidel:2018apz}. Metaparticles arise as zero modes of metastrings propagating on modular (doubled) spacetime and obey a modified dispersion relation exhibiting intrinsic UV/IR mixing controlled by a duality scale $\mu$. Using a generalized Bekenstein argument adapted to metaparticles, we derive quantum-corrected entropy contributions associated with geometric and dual (winding-like) sectors of the underlying phase space. 

When treated independently, these two entropy branches lead to an incomplete thermodynamic description, exhibiting unphysical behavior at small horizon area. We show that consistently treating the metaparticle as a single entangled quantum object—rather than as two independent sectors—naturally resolves these pathologies. We propose a pseudo-entangled total entropy that incorporates correlations between the geometric and dual sectors. The reality requirement of the entropy dynamically enforces a minimal horizon area and, equivalently, a minimal effective length scale associated with modular spacetime.

The resulting black hole thermodynamics exhibits a finite maximal temperature, a divergence of the heat capacity signaling a continuous phase transition, and a shutdown of Hawking radiation through geometric channels, leaving behind a cold, stable remnant. Unlike matter-supported or curvature-bounded regular black holes, the remnant obtained here is non-material and non-geometric in nature, corresponding to a finite modular core of spacetime rather than a dust-filled interior. We compare this scenario with mimetic gravity \cite{ChamseddineMukhanov2013} and other non-singular black hole models, emphasizing the distinct role played by first-class constraints, entropy, and modular geometry in the present framework.
\end{abstract}

\pagebreak

\tableofcontents

\pagebreak


\section{Introduction}
Several approaches to quantum gravity have indicated departures from the regular relativistic dispersion relation which came to be known as modified dispersion relations (MDR's for short), as well as a need to generalize Heisenberg's uncertainty principle (GUP). It was argued in \cite{Chen:2002tu} that when gravitational effects were to be included, the GUP could prevent black holes from totally evaporating in the same manner that the Heisenberg uncertainty principle prevents the Hydrogen atom from total collapse. As shown in \cite{Hossenfelder:2005ed}, any adequate theory of quantum gravity predicting a minimal length must contain three ingredients: a form of deformed special relativity (DSR)\cite{Amelino-Camelia:1999hpv}, MDR's and GUP's. A direct link between these three were shown to exit by the same author\footnote{In future work\cite{Chouha-PR-cosm:2026-c}, we will show how an extended generalized uncertainty principle (EGUP) can be obtained from the metaparticle we introduce below}.
Lacking a final theory of quantum gravity and without direct experimental guidance, one has to go beyond effective field theories and study the effects of the new ingredients that one finds in the different quantum gravity proposals hoping to get a hint at the right path to follow. More likely than not, the final quantum gravity theory (if such a theory exists) will contain ingredients from the various different approaches proposed so far. 
\par There are two natural testing grounds for any inspired quantum gravity theory, namely, early universe cosmology and black hole physics. On the cosmological side, the origin of this work was an interest in describing the dynamics of string gas cosmology (SGC)\cite{Brandenberger:2008nx}. Although Double Field Theory (DFT) \cite{Hull:2009mi} captures some of the dynamics of SGC, it does not contain enough stringy effects since it is missing the oscillatory modes of the strings. In addition, due to a technical requirement of the theory, the section condition\footnote{The requirement that the fields and their derivatives are independent of the dual-coordinates}, it does not offer a proper description of the dynamics of the Hagedorn phase where both the winding and momentum modes coexist. Therefore, one wonders whether there is another formulation of string theory in which unexplored "{\it{stringy}}" features are captured and can have cosmological 
consequences \cite{Chouha-PR-cosm:2026} as well as other phenomenological effects on black hole physics such as those we start to explore in this work. 
\par One such approach is known as {\bf{Metastrings}} \cite{Freidel:2015pka} which we introduce in appendix \ref{MString}.\ As we summarize in the appendices, starting from a new physical principle, and its mathematical realization leads naturally  to this metastring formulation. In brief, metastring theory is a manifestly (non-commutative) T-dual covariant (phase space) formulation of string theory that includes a new mathematical ingredient, the non-flat\footnote{By non-flat we mean that it is not just given by the constant $\omega$ in equation \ref{Eq:Eta-Omega-H-zeroB}.} symplectic two-form $\omega$ defined by (\ref{Eq:Eta-Omega-H-B}). (hence the "meta" in metastring since it is not defined using the usual Polyakov string action").
\par Metastrings arise when one tries to construct a quantum spacetime  reconciling the existence of a fundamental natural length such as the Planck length or the string length with Lorentz invariance. This new notion of quantum spacetime is called {\it{modular spacetime}} \cite{Freidel:2017xsi} which captures the idea of Aharanov \cite{Aharonov:1969qx}, that the only purely quantum observables are modular in nature, as explained in appendix \ref{ModST}. The zero modes of the metastrings are the metaparticles (c.f. \ref{Metapart}) introduced in \cite{Freidel:2018apz} which obey a new modified dispersion relation (\ref{Eq:MDR}) that we extensively use in this work.
\par Before we proceed, we want to emphasize that as has been remarked in \cite{Murk:2023rwl}, throughout our study, we will only consider the Schwarzschild black hole since we are interested in the end-life of such objects and as has been shown by Page in \cite{Page:1976df} during the Hawking process, angular momentum is shed much faster than mass and as shown by Gibbons \cite{Gibbons:1975kk} charged black holes rapidly discharge in Schwinger-like pair production processes \cite{Schwinger:1951nm}.\footnote{We do not discuss the extremal case (c.f. \cite{Gralla:2025gzl} for a recent interesting analysis).} 

\par This work is divided as follows: in section \ref{Reg-BH} we review some of the known results on regular (non-singular) black holes, then, in section \ref{sec:Metapart-Metastring} we introduce metastrings and metaparticles studying in detail the modified dispersion relation arising form metaparticles which will be the focal point of our current work on black hole thermodynamics. In section \ref{Sec:Meta-BH} we perform a detailed analysis of the metaparticle modified black hole thermodynamics. Finally in section \ref{Sec:Disc-Comparison} we end our study with a detailed discussion of the modified evaporation scenario of the metaparticle corrected black hole comparing it to other approaches and in particular to mimetic gravity's non-singular black holes\cite{ChamseddineMukhanov2017LC}. In order not to disrupt the flow of the paper, we relegate detailed and more technical discussions to the appendices. In Appendix \ref{SubSec:Original-Bek} we review the original Bekenstein argument; in Appendix \ref{ModST} we introduce quantum modular spacetime followed by a review of metastrings in Appendix \ref{MString}. We conclude our appendices by a discussion of the validity of the adiabatic treatment of this work in Appendix \ref{App:Adiabaticity}.

\section{Regular Black Holes and Black Hole Mimickers}\label{Reg-BH}
Historically, the first non-singular black hole was proposed by Bardeen in 1968 \cite{Bardeen1968} followed by Gliner's work \cite{gliner1969regular}.
Both works either modified the black hole metric or the energy-momentum tensor in order to evade the Penrose-Hawking singularity theorems \cite{hawking1969singularity}. 
 \par More recently, many works tried to write down effective metrics based on extra ingredients coming from different approaches to quantum gravity. On the sting-theoretic side, such works as \cite{Nicolini2019} based on results linking Toroidal duality (T-duality) and the existence of a minimal length\footnote{ Taken to be the string length $l_{s}$.}\cite{Padmanabhan:1996ap,Smailagic:2003hm}, the authors derived modifications to the Newtonian potential thus obtaining a modified mass function giving rise to an effective spherically symmetric metric incorporating the minimal-length. The authors found a departure from the regular behavior of the Hawking temperature \cite{Hawking1974a,Hawking1975}, in particular, they found that as the black hole evaporates, its temperature reaches a maximum then decreases to zero when the minimal mass is reached, in sharp contrast to the diverging Hawking temperature. Further, in \cite{Jusufi:2024dtr} the authors showed that the geodesics are complete in such an effective spacetime metric suggesting that on short scales gravity changes its character and becomes repulsive producing a screening effect such as that observed in the Debye screening in Electrostatics \cite{DebyeHuckel1923}. 
We also note an earlier different approach 
using a Generalized Uncertainty Principle (GUP). The gravitationally corrected GUP appeared in many works such as \cite{Adler:2001vs,Adler:2010wf, Scardigli:1999jh, Chen:2002tu, Nozari:2006bi} investigating corrections to the Hawking temperature and other thermodynamical functions. Those works improved on the original Hawking diverging temperature and found that the temperature reached a maximum at the Planck mass in a finite Shwarzschild time and suggested the possibility of having a black hole remnant. However, we must note that the specific heat goes to zero at a finite maximal temperature corresponding to the Planck mass, this anomalous behavior indicates an instability of the remnant. Similar conclusions were obtained using the Black Hole Uncertainty Correspondence (BHUP) \cite{Carr:2014mya} which states that there could exist a connection between the UV and IR regimes, that is between the Compton wavelength (on microscopical scales) and the black hole's Shwarzschild radius (on macroscopic scales). 
By combining the two lengths, either linearly (obtaining an expression similar to that of the GUP) or quadratically, one can get corrections to the Shwarzschild radius for $M>M_P$ 
The study proposed
the possibility of forming sub-Planckian black holes with the size of the order of their Compton wavelength.
\par An alternative to regular black holes are black hole mimickers first appearing in \cite{Lemos:2008cv} and reviewed in \cite{Carballo-Rubio:2023mvr,Carr:2015nqa}. These are horizonless spacetimes that do not have a trapped region\footnote{A trapped region on a 3-dimensional hypersurface is the set of all the points on the hypersurface through which a trapped surface passes. The latter is a closed, spacelike, 2-surface such that both expansion of ingoing and outgoing null geodesics are negative \cite{Carballo-Rubio:2025fnc}.} and can be astrophysical objects such as ultracompact stars or even more speculatively traversable wormholes. The absence of  an event horizon (or any horizon for that matter) opens up observational possibilities given that their interiors are not causally disconnected from the exterior. Thus, this opens up the possibility of electromagnetic waves propagating and interacting with their interior resulting in new photon rings appearing in the black hole images  \cite{Eichhorn:2022bbn,Mazur:2015kia,Carballo-Rubio:2022aed,EventHorizonTelescope:2022xqj} or to observable thermal emissions due to the heating up of both the interior and exterior of such objects. The possibility of gravitational waves interacting with the interior of such objects has been proposed in \cite{Cardoso:2016rao,Cardoso:2017cqb} resulting in gravitational wave echoes. \cite{EventHorizonTelescope:2022xqj,Carballo-Rubio:2023fjj}.    
\par The above discussion shows that we are in a time where some of the imprints and predictions of the different approaches to quantum gravity are within our phenomenological grasp and can have observational consequences. The purpose of this work is to investigate further what Metastrings and metaparticles can tell us about the thermodynamical evolution of black holes which serves two purposes: the first is to better understand the physical parameters appearing in the theory, namely, the dualities, the physical realization of a dual geometry, the possible role of fluxes in the final stages of a black hole's evolution, etc. While the second is to investigate what this new reformulation of string theory can teach us about black holes: what different thermal history the black hole would have if it absorbs a metaparticle as opposed to regular Hawking-like radiating pairs? Does the Hawking radiation halts? Is the remnant stable? What is the nature of this remnant \cite{Chouha-PR-cosm:2026-b}? What is the effective-metric\footnote{More precisely, we mean the class of effective-metrics, as the metric is not unique as discussed in \cite{Chouha-PR:2025}.} describing the metaparticle induced black hole scenario \cite{Chouha-PR:2025}?  
\par Since our main study focuses on the thermodynamical evolution of black holes\footnote{In Appendix \ref{App:Adiabaticity} we show that our adiabatic treatment is justified throughout the thermal evolution of the metaparticle-modified black hole.}, we are going to approach the problem by a well established general relativistic result due to Bekenstein in order to study the modifications to the area of the black hole's event horizon as it absorbs a pure quantum object such as the metaparticle (as explained in the text below).
The Bekenstein argument is derived in Appendix \ref{SubSec:Original-Bek} which also serves to establish notation and conventions as well as the physics of the gedanken experiment. We will see that the conclusions that we can draw are mainly due to this quantum nature of the metaparticle and in particular to the UV/IR-mixing that is naturally built in this formalism. 
\section{Metastrings and Metaparticles: New Horizons}\label{sec:Metapart-Metastring}
In this section we briefly outline the new concepts of metaparticles \cite{Freidel:2018apz} as arising from metastrings\cite{Freidel:2015pka, Freidel:2017xsi} and the novelties they introduce, and then proceed to study the effect of metaparticles on black hole thermodynamics. For more in-depth insight, the reader is referred to the appendices at the end of this work. \\

\subsection{Metaparticles}\label{Metapart}
In the metastring/metaparticle framework spacetime is intrinsically doubled,
consisting of geometric coordinates $x^\mu$ and dual coordinates $\tilde{x}_\mu$.
Quantization of the underlying phase-space structure implies the fundamental
non-commutative relation
\begin{equation}\label{Eq:xxcomm}
[x^\mu,\tilde{x}_\nu]=2\pi i \lambda^2\delta^\mu_{\nu},
\end{equation}
which characterizes the modular (quantum) spacetime on which metaparticles
propagate (see Appendix~\ref{ModST} for details).

Metaparticles can be thought of as the generalization of particle excitations that
appears in metastring theory. In particular they represent the zero modes of the metastrings\cite{Freidel:2018apz}. Their dynamics is described by the worldline action
\begin{equation}\label{Eq:Metaparticle}
S_{MP}=\int d\tau \left[p\cdot\dot{x}
+\tilde{p}\cdot\dot{\tilde{x}}
+\pi\lambda^2 p\cdot\dot{\tilde{p}}
-\frac{N}{2}(p^2+\tilde{p}^2+2\mathfrak{m}^2)
+\tilde{N}(p\cdot\tilde{p}-\mu)\right].
\end{equation}

Here $x$ and $\tilde{x}$ denote the spacetime and dual-spacetime coordinates,
with conjugate momenta $p$ and $\tilde{p}$ respectively.
The Lagrange multipliers $N$ and $\tilde N$ enforce the two first-class
constraints of the theory: $N$ imposes the generalized mass-shell (Hamiltonian)
constraint, while $\tilde N$ enforces the duality (diffeomorphism) constraint
$p\cdot\tilde{p}=\mu$, which couples the geometric and dual momentum sectors.
This relation implies that for a metaparticle of energy $E$ there exists a dual
excitation with energy $\tilde{E}=\mu/E$, reflecting the non-factorizing nature
of the metaparticle as a single quantum object composed of mutually entangled
particle and dual degrees of freedom. In this sense, a single local geometric description of spacetime is incomplete\footnote{In the present framework spacetime is not fundamental but arises from a choice of polarization of modular (doubled) phase space. Different polarizations related by $O(d,d)$ transformations describe the same underlying physical state, so geometry is polarization-dependent rather than fundamental. Locality therefore emerges only in regimes where the dual sector can be consistently suppressed; when this suppression fails, no single geometric description captures the full constraint structure. The resulting non-locality is modular in nature and does not imply a violation of infrared microcausality.}.

The phase-space origin of the symplectic term
$\pi\lambda^2 p\cdot\dot{\tilde{p}}$ and the doubled field description are
summarized in Appendices~\ref{ModST} and~\ref{MString}.
\par After choosing a particular polarization of the phase space, i.e. states labeled for ex. by $\ket{p,\tilde{p}}$, the metaparticle propagator \cite{Freidel:2018apz} between off-shell states can be computed from the wordline of the action (\ref{Eq:Metaparticle}) and is given by:
\begin{equation}\label{Eq:Propagator}
    G(p,\tilde{p})\sim \frac{\delta(p\cdot\tilde{p}-\mu)}{p^2+\tilde{p}^2+2\mathfrak{m}^2-i\epsilon}.
\end{equation}
\par Thus, a generic metaparticle is characterized by three important invariant scales, each associated with a particular mathematical structure of the construction of the quantum spacetime and metastrings outlined in the appendices \ref{ModST} and \ref{MString} respectively: the fundamental non-commutative scale $\lambda$, the metaparticle "mass" scale $\mathfrak{m}$ and the duality scale $\mu$.
\par If we go to a particular gauge in which for example $\tilde{p}$ is a space-like vector, we get a {\bf{modified dispersion relation}} (MDR):
\begin{equation}\label{Eq:MDR}
    E^2+\frac{\mu^2}{E^2}={\vec{p}}^{\ 2}+2\mathfrak{m}^2
\end{equation}
which in view of the duality scale $\mu$ mixes the UV/IR scales which is one the very attractive features of metaparticles for cosmology \cite{Chouha-PR-cosm:2026} (c.f. \cite{Freidel:2021wpl} for some application to dark matter/energy) and for black hole thermodynamics which we explore in this work. 
\par In this gauge-fixed description, $E$ denotes the physical energy conjugate to the
chosen time variable.

\subsection{A Closer Look at the Metaparticle Dispersion Relation}\label{Met-MDR}
Since we assume that during the Hawking evaporation process only massless particles are emitted, we will consider the special massless case of the modified dispersion relation (\ref{Eq:MDR}). As analyzed in \cite{Freidel:2021wpl} the massless case corresponds to the equality of the mass parameter with the duality parameter i.e. $\mathfrak{m}=\mu$ respectively.\\
Thus the metaparticle  MDR (\ref{Eq:MDR})
will read
\begin{equation}\label{Eq:MDR-Massless}
    E^2+\frac{\tmu^2}{E^2}={\vec{p}}^{\ 2}+2\tmu
\end{equation}
where all quantities are written in terms of their corresponding Planckian values. We note that the duality parameter $\mu$ has dimensions of energy squared\footnote{Sometimes, we write $\tilde{\mu}$ instead of $\mu$ to remind the reader that it is the dimensionless quantity in units of $E_{\text{P}\,}^2$.}.
The spectrum is invariant under two duality symmetries:
\begin{equation}\label{Eq:Duality-Sym}
E\leftrightarrow-E,\hspace{3mm}E\leftrightarrow\frac{\tmu}{E}.     
\end{equation}
The first symmetry means that to each particle is associated an anti-particle, while the second symmetry means that to each of those pairs is associated a dual that is a dual-particle and anti-dual-particle. This is also manifest from the dispersion relation, which has four solutions, each corresponding to a particular particle.\footnote{\label{fn:energy}
Equation~(\ref{Eq:Solutions}) arises from solving the constraints of the reparametrization-invariant
metaparticle theory after gauge fixing (or, equivalently, choosing a polarization).
Prior to gauge fixing, the energy variable parametrizes the constraint surface and
does not yet represent a physical Hamiltonian. Once a physical time is chosen and the
theory is reduced to its physical phase space, the same quantity coincides with the
physical energy conjugate to that time.
The two algebraic branches correspond to metaparticle and anti-metaparticle sectors
(or, equivalently, opposite time orientations), each admitting a Hamiltonian bounded
from below; the presence of negative-energy solutions therefore does not imply an
instability or an energy spectrum unbounded from below.
This structure is directly analogous to the treatment of the relativistic particle
and to the implementation of the Virasoro constraints in string theory, where
reparametrization invariance leads to multiple branches of solutions while the
physical spectrum in each sector remains bounded below
(see, e.g., \cite{Polchinski:1998rq,Green:1987sp}).
}

\begin{equation}\label{Eq:Solutions}
\begin{split}
    & E^{++}=\frac{p+\sqrt{p^2+4\tmu}}{2}, \quad E^{+-}=\frac{p-\sqrt{p^2+4\tmu}}{2},\\
    &E^{-+}=\frac{-p+\sqrt{p^2+4\tmu}} {2},\quad E^{--}=\frac{-p-\sqrt{p^2+4\tmu}}{2}.
\end{split}    
\end{equation}
We see from the above that $E^{++}=-E^{--}$ so $E^{--}$ is the antiparticle of $E^{++}$. By taking the conjugate, we can write $E^{++}=\dfrac{\tilde{\mu}}{E^{-+}}=-\dfrac{\tmu}{E^{+-}}$ and hence, $E^{-+}$ is the dual particle of $E^{++}$, while $E^{+-}$ is the anti dual-particle of $E^{++}$. 
\begin{equation}\label{Eq:Sol.Meta-anti}
    (E^\pm)^2=\frac{p^2+2\tmu\,\pm\sqrt{p^4+4\tmu\,p^2}}{2}
\end{equation}\\
It is important to note that a metaparticle is a single quantum object that is entangled with its dual. We can write pictorially 

\begin{equation}\label{Eq:Ket-meta}
\ket{\text{Metaprticle}}_{\text{Full}}\sim\ket{E}+\ket{\tE}.
\end{equation}
Where by $\ket{E}$ and $\ket{\tE}$ we denote the quantum state of the particle and its dual respectively.
This entanglement plays an important role in our study below. \\
\par In order to get a physical feeling of what metaparticles are, we have recourse to cosmology. Assuming a homogeneous and isotropic cosmological background described by the Friedmann--Lema{\^i}tre--Robertson--Walker (FLRW) metric \cite{Friedmann:1922,Friedmann:1924,Lemaitre:1927,Robertson:1935,Walker:1937}, the continuity equation (expressing the conservation of the energy-density) is given by
\begin{equation}\label{Eq:FRW-Cont}
\dot{\epsilon}+3H(\epsilon+P)=0    
\end{equation}
where $H$ denotes the Hubble parameter $H\equiv\dot{a}(t)/a(t)$, $a(t)$ is the scale factor and a dot denotes a derivative with respect to cosmic-time $t$. $\epsilon$ denotes the total energy density of a gas (say) of metaparticles that we can write as $\epsilon=n\,E$ where $n$ is the number density of particles and $E$ is the energy per metaparticle. $P$ is the total pressure that we can relate to the energy density via the equation of state parameter $w$, $P=w\, \epsilon$.\\
The value of $w$ tells us a lot about the nature of the gas that is present at that epoch in the history of our Universe. For example, $w=0$ denotes dust or pressure-less matter, while $w=1/3$ denotes a relativistic gas such as radiation. The scale factor (which gives us information about the cosmic evolution of the Universe) will have different time dependence according to whether we are in a radiation-dominated or matter-dominated era. Solving for $w$, we get
\begin{equation}\label{Eq:w-param}
    w=-\frac{\dot{E}}{3HE}.
\end{equation}
Let us compute $w$ for the metaparticle dispersion relation $E^{++}$.
\begin{equation}\label{Eq:MDR-Meta-phys}
    E^{++}_{\text{phys}}=\frac{p_{\text{phys}}+\sqrt{p_\text{phys}^2+4\tmu}}{2}
\end{equation}
where in an expanding universe, we need to consider physical quantities i.e. those that we measure in a local frame at time $t$. The physical momentum $p_{\text{phys}}$ is related to the comoving\footnote{Comoving coordinates are glued on the expanding (or contracting) space and their evolution in time is due to the scale factor $a(t)$. Here we assume that $p_{\text{com}}\neq0$.} momentum via the scale factor $p_\text{phys}=\dfrac{p_\text{com}}{a(t)}$ and therefore, $\dot{p}_\text{phys}=-\dfrac{p_\text{com}\dot{a}(t)}{a(t)^2}=-p_\text{phys}\, H$.\\
Computing $\dot{E}^{++}_{\text{phys}}$ we get
\begin{equation}\label{Eq:dot-E-phys}
  \dot{E}^{++}_{\text{phys}}=\frac{\dot{p}_\text{phys}}{\sqrt{p_\text{phys}^2+4\tmu}} E^{++}_{\text{phys}}=-  \frac{p_\text{phys}H}{\sqrt{p_\text{phys}^2+4\tmu}} E^{++}_{\text{phys}}, 
\end{equation}
and therefore the equation of state parameter (\ref{Eq:w-param}) reads
\begin{equation}\label{Eq:w-param-metapart}
    w_\text{metap}=\frac{p_\text{com}}{3a(t)\sqrt{\frac{p_\text{com}^2}{a(t)^2}+4\tmu}}=\frac{p_\text{com}}{3\sqrt{p_\text{com}^2+4\tmu\, a(t)^2}}.
\end{equation}
Immediately, we see that in the early Universe as $a(t)\to0$ (i.e. also at high energies and hence temperatures) $w_\text{metap}\to\dfrac{1}{3}$. That is, metaparticles behave as relativistic matter or radiation. Whereas, at late times i.e. $a(t)\to\infty$, $w_\text{metap}\to0$ the metaparticles behave as dust or pressure-less matter. 
\par As for the dual metaparticle $E^{-+}$, the equation of state parameter will be given by
\begin{equation}\label{Eq:w-param-dual-metapart}
    w_\text{dual-metap}=-\frac{p_\text{com}}{3a(t)\sqrt{\frac{p_\text{com}^2}{a(t)^2}+4\tmu}}=-\frac{p_\text{com}}{3\sqrt{p_\text{com}^2+4\tmu\, a(t)^2}}.
\end{equation}
and hence in the early Universe as $a(t)\to0$, $w_\text{dual-metap}\to-\dfrac{1}{3}$ which through a string-theoretic interpretation is nothing but winding modes; while at late times $a(t)\to\infty$, $w_\text{dual-metap}\to0$ behave as pressure-less dust. Thus, as in string gas cosmology, where winding modes play a very important role during the Hagedorn phase, we expect dual metaparticles to play an important role as the black hole starts to evaporate and its temperature rises.
\par Having gained some insight into the nature of metaparticles, we now turn to their role in black hole thermodynamics and the possible fate of black hole evaporation.

\section{Metaparticles and Black Hole Thermodynamics}\label{Sec:Meta-BH}
We follow the Bekenstein Argument \cite{Bekenstein:1973ur,Bekenstein:1974ax} and Geroch process \cite{https://doi.org/10.1111/j.1749-6632.1973.tb41445.x} outlined in Appendix \ref{SubSec:Original-Bek} except that now the quantum matter that the black hole absorbs is a metaparticle. Keeping in mind that it is an entangled quantum object, we will proceed in two steps: we first ignore the effects of quantum entanglement between the metaparticle and its dual in order to isolate the physical consequences of the metaparticle's modified dispersion relation from those due to entanglement. We then include the quantum entanglement and compare the physical effects of both on the black hole evaporation process and its final fate.   
\subsection{Bekenstein's Argument in Light of Metaparticles Without Entanglement}\label{subsec:Bek-Meta}
Without repeating the full arguments of appendix \ref{SubSec:Original-Bek} we still assume that the regular Heisenberg uncertainty relation holds $\delta p\,\delta x\geq1/2$ but we now have a modified dispersion relation.
Usually one proceeds by relating the uncertainty in momentum $\delta p$ to that in energy $\delta E$ via the relativistic dispersion relation, but here we are going to use the metaparticle's MDR, equation (\ref{Eq:MDR-Massless}) to get
\begin{equation}\label{Eq:DeltaE-Deltap-Meta}
    \delta p=\delta E \left(1+\frac{\tmu}{E^2}\right)
\end{equation}
To compute the uncertainty in momentum $\delta p$ (as a function of $\delta E$), due to the metaparticle we can either plug the expression for $E^{++}$ (\ref{Eq:Solutions}) or use the solution $E^{+\,2}$ from (\ref{Eq:Sol.Meta-anti}) into (\ref{Eq:DeltaE-Deltap-Meta}). In both cases we get a modified uncertainty-relation between energy and position given by:
\begin{equation}\label{Eq:DeltaEPlus-Deltax}
    \delta E^{+}\,\delta x\geq\frac{p^2+4\tmu +p\sqrt{p^2+4\tmu}}{4(p^2+4\tmu)},
\end{equation} and we see that as $\tmu\to 0$, we recover $\delta E\,\delta x\geq1/2$.\\
Plugging this result in (\ref{Eq:Area-Plnakc-units}), and assuming that the fluctuations in momentum are of the same order of $p$ i.e. $\delta p\sim p$ and using the unmodified Heseinberg uncertainty relation $\delta p\,\delta x\geq 1/2$, we can write the minimal increase in area for absorbing a metaparticle-anti-metaparticle (pair) as
\begin{equation}\label{Eq:Delta_Area-Meta}
    (\Delta A)^+_{\text{min}}=
    2\pi\,\Gamma\,\frac{\left(1+\sqrt{1+16\tmu\,\delta x^2}\right)}{\sqrt{1+16\tmu\,\delta x^2}}.
\end{equation}
For consistency, recalling that classically $\Gamma\equiv 1$, and that $A$ is in units of $L_P^2$ we get that the minimal increase in area as $\tmu\to0$ is $4\pi
\,L_P^2$. Thus, the area of a black hole increases by units of the Planck area, lending support to the picture that the event horizon (or more generally the trapped surface) of the black hole is quantized in units of the Planck area.

\par The minimal increase in entropy accompanying the minimal increase in the surface area of the black hole is $\ln{2}$ corresponding to 1 bit of information  (c.f. the discussion in appendix \ref{SubSec:Original-Bek}). And thus we write
\begin{equation}\label{Eq:MetaPos-dS/dA}
\frac{(\Delta S^{+})_{\text{min}}}{(\Delta A)_{\text{min}}}\simeq \frac{dS^{+}}{dA}=\frac{\ln2}{2\pi\,\Gamma}\left(1+\frac{\pi}{4\tmu A}-\frac{\pi}{4\tmu A}\sqrt{1+\frac{4\tmu A}{\pi}}\right),
\end{equation}
where we identified $\delta x\equiv r_H=2\tM$ (where $\tM\equiv\dfrac{M}{M_P}$ is the mass in units of the Planck mass $M_P$) and $A\equiv 4\pi\, r_H^2$.Upon integration over the area we get the contribution of the metaparticle-anti-metaparticle pair absorption to the entropy $S^{+}$ of the black hole
\begin{equation}\label{Eq:Entropy-Meta+}
    S^{+}(A)=\frac{\ln2}{2\pi\,\Gamma}\left[A+\frac{\pi}{2\tmu}\ln\left(\frac{4\tmu\,A}{\pi}\right)-\frac{\pi}{2\tmu }\ln\left(1+\sqrt{1+\frac{4\tmu A}{\pi}}\right)-\frac{\pi}{2\tmu }\sqrt{1+\frac{4\tmu A}{\pi}}\,\right]
\end{equation}
We now need the contribution of the dual-metaparticle and anti-dual metaparticle pair to the entropy. Proceeding as above, we relate $\delta E^{-}$ to $\delta p$ using either $E^{-+}$ or $E^{-\,2}$ in (\ref{Eq:Solutions}) and we get a modified uncertainty relation
\begin{equation}\label{Eq:DeltaEMinus-Deltax}
    \delta E^{-}\,\delta x\geq\frac{p^2+4\tmu -p\sqrt{p^2+4\tmu}}{4(p^2+4\tmu)},
\end{equation} and this time we see that as $\tmu\to 0$, we have $\delta E^{-}\,\delta x\geq 0$ i.e. the result is trivially satisfied since when the duality parameter $\tmu$ vanishes there is no dual-metaparticle present (or seen another way when $\tmu=0$ the dual spacetime decouples from spacetime and we do not have access to the physical duals anymore).\\
The minimal increase in area accompanying the absorption of the dual-anti-dual metaparticle pairs will be
\begin{equation}\label{Eq:Delta_Area-Meta}
    (\Delta A)^-_{\text{min}}=
    2\pi\,\Gamma\,\frac{\left(\sqrt{1+16\tmu\,\delta x^2}-1\right)}{\sqrt{1+16\tmu\,\delta x^2}},
\end{equation}
    and once again we see that when $\tmu\to0$, there is no change in the surface area coming from the dual sector. 
    \par The associated increase in entropy $S^{-}$ is given by
\begin{equation}\label{Eq:Entropy-Meta-}
    S^{-}(A)=\frac{\ln2}{2\pi\,\Gamma}\left[A+\frac{\pi}{2\tmu }\ln\left(1+\sqrt{1+\frac{4\tmu A}{\pi}}\right)+\frac{\pi}{2\tmu }\sqrt{1+\frac{4\tmu A}{\pi}}\,\right]
\end{equation}    
Now, adding the two entropies linearly (ignoring entanglement), and fixing the calibration factor $\Gamma=\dfrac{4\ln2}{\pi}$ (to recover Hawking's leading term) we finally get:
\begin{equation}\label{Eq:STot-noEntg}
\boxed{S_{\text{Tot}}(A)=\frac{1}{4}\left[A+\frac{\pi}{4\tmu}\ln\left(\frac{4\tmu\,A}{\pi}\right)\,\right].}    
\end{equation}
The first observation is that we recover the well-known logarithmic correction to
black-hole entropy, which arises generically in a wide variety of quantum-gravity
approaches, including loop quantum gravity, entanglement entropy calculations,
string-theoretic analyses, and one-loop quantum corrections
\cite{KaulMajumdar2000,Solodukhin2011,Sen2013Log,Fursaev1995}. 
The sign of the logarithmic coefficient is not universal: loop quantum gravity
calculations typically yield a negative coefficient, while entanglement and
one-loop quantum corrections produce model-dependent coefficients that may be
either positive or negative. In the present framework the logarithmic term
appears with a positive coefficient.

\begin{figure}[h!]
    \centering
    \includegraphics[width=0.8\textwidth]{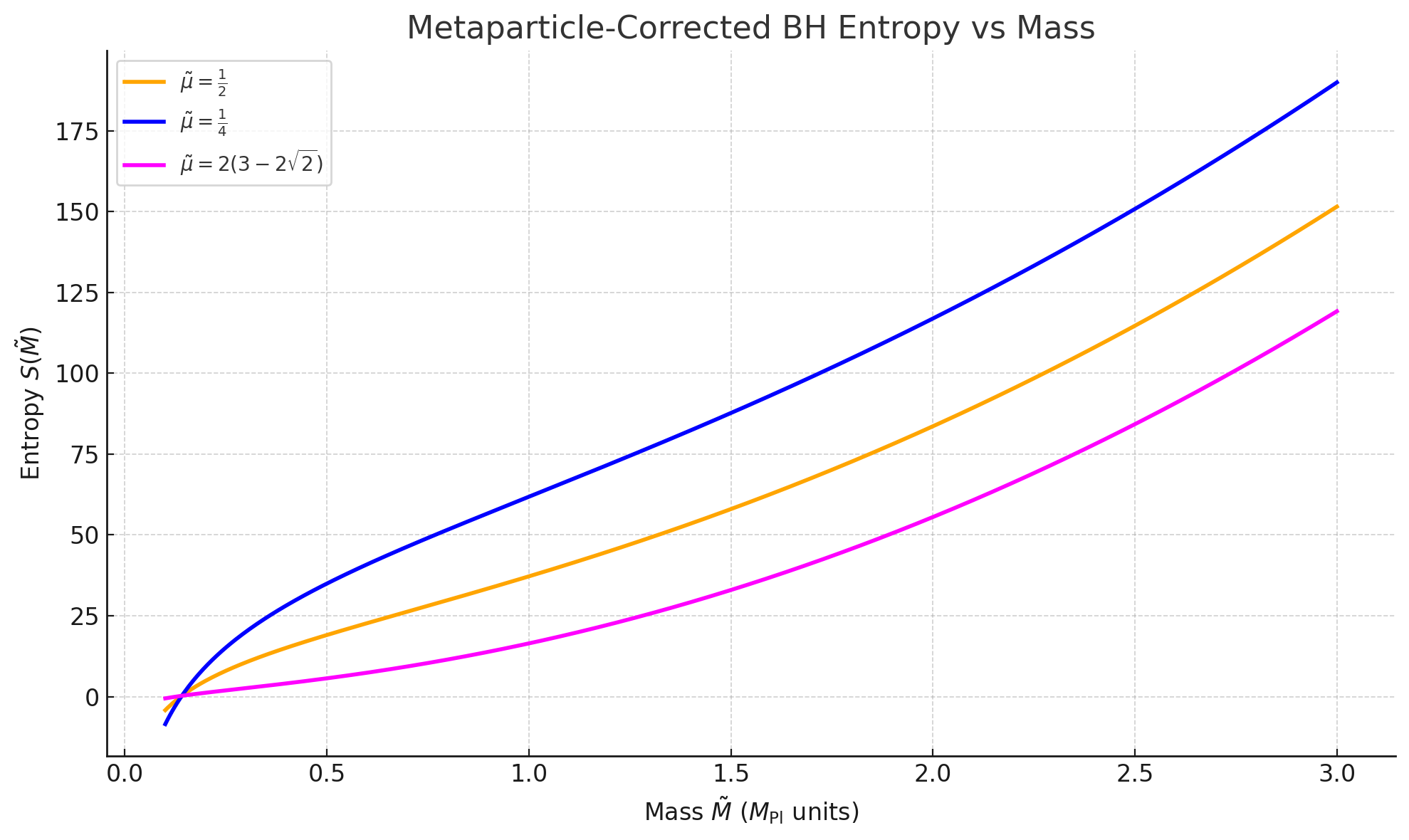}
    \caption{
        Metaparticle-corrected black hole entropy as a function of mass $\tilde{M}$ in Planck units for three values of the metaparticle- duality parameter $\tilde{\mu}$. All curves include logarithmic corrections to the standard Bekenstein-Hawking entropy. We particularly show the functional behavior for small masses.                   }
    \label{fig:metaparticle_entropy}
\end{figure}

As can be observed from figure \ref{fig:metaparticle_entropy}, at large masses, all models asymptotically approach the classical area law. However, for $\tilde{M} \lesssim 1$, the entropy deviates significantly due to quantum corrections (resulting from the duality pairing between the metaparticle and their dual). These deviations are stronger for smaller $\tilde{\mu}$\footnote{The three values of $\tmu$ quoted in the plots come from identifying the maximal temperature with the Hagedorn temperature in different flavors of string theory: Type II, Bosonic and Heterotic respectively \cite{Chouha-PR-cosm:2026-b}.} and play a central role in modifying the black hole's evaporation dynamics.

\par The temperature (in Planck units) can be computed from the thermodynamical relation 
\begin{equation}\label{Eq:TdM}
    \tT=\frac{\partial\tM}{\partial\tS}
    \end{equation}
    and we get (\ref{Eq:TdM}) 
\begin{equation}\label{Eq:Temp-Meta-No-int}
    \tT(\tM)=\frac{8\tmu\,\tM}{\pi\left(1+64\tmu\,\tM^2\right)}
\end{equation}
which reaches a maximum value
\begin{equation}\label{Eq:T_max-Meta-No-ent}
\tT_{\text{max}}=\frac{\sqrt{\tmu}}{2\pi},
\end{equation}
when the black hole reaches a critical mass
\begin{equation}\label{Eq:Critical-mass-no-Ent}
\tM_{\text{crit}}=\dfrac{1}{8\sqrt{\tmu}}.
\end{equation}
Furthermore, to complete our understanding of the black hole's thermodynamical evolution and stability, we need to compute its heat capacity $C$ via
\begin{equation}\label{Eq:Heat_Cap}
    C=\frac{\partial\tM}{\partial\tT}
\end{equation}
and we find

\begin{equation}\label{Eq:Heat-Cap-Meta-No-Ent}
    C(\tM)=\frac{8\tmu}{\pi}\,\frac{(1+64\tmu\,\tM^2)^2}{1-64\tmu\,\tM^2}.
\end{equation}
In what follows,  we will write the temperature $\tT$, the mass $\tM$ and the heat capacity $C$ as rescaled variables $\ttau$, $\tm$ and $\tC$ independent of $\tmu$, such that 
\begin{equation}\label{Eq:rescaled-vars}
\ttau\equiv\tT/\sqrt{\tmu},\qquad \tm\equiv\sqrt{\tmu}\,\tM, \quad\text{and}\quad \tC=C/\tmu.
\end{equation}
The reason is that we want our physical discussion of the modified metaparticle black hole to be independent of the possible different values that the duality parameter $\tmu$ may take. 
\par We plot the behavior of the rescaled temperature $\ttau$ and rescaled heat capacity $\tC$ in figure \ref{Fig:Meta-Temp-Heat-Capacity} below.
\begin{figure}[h!]
\centering
\includegraphics[width=0.48\textwidth]{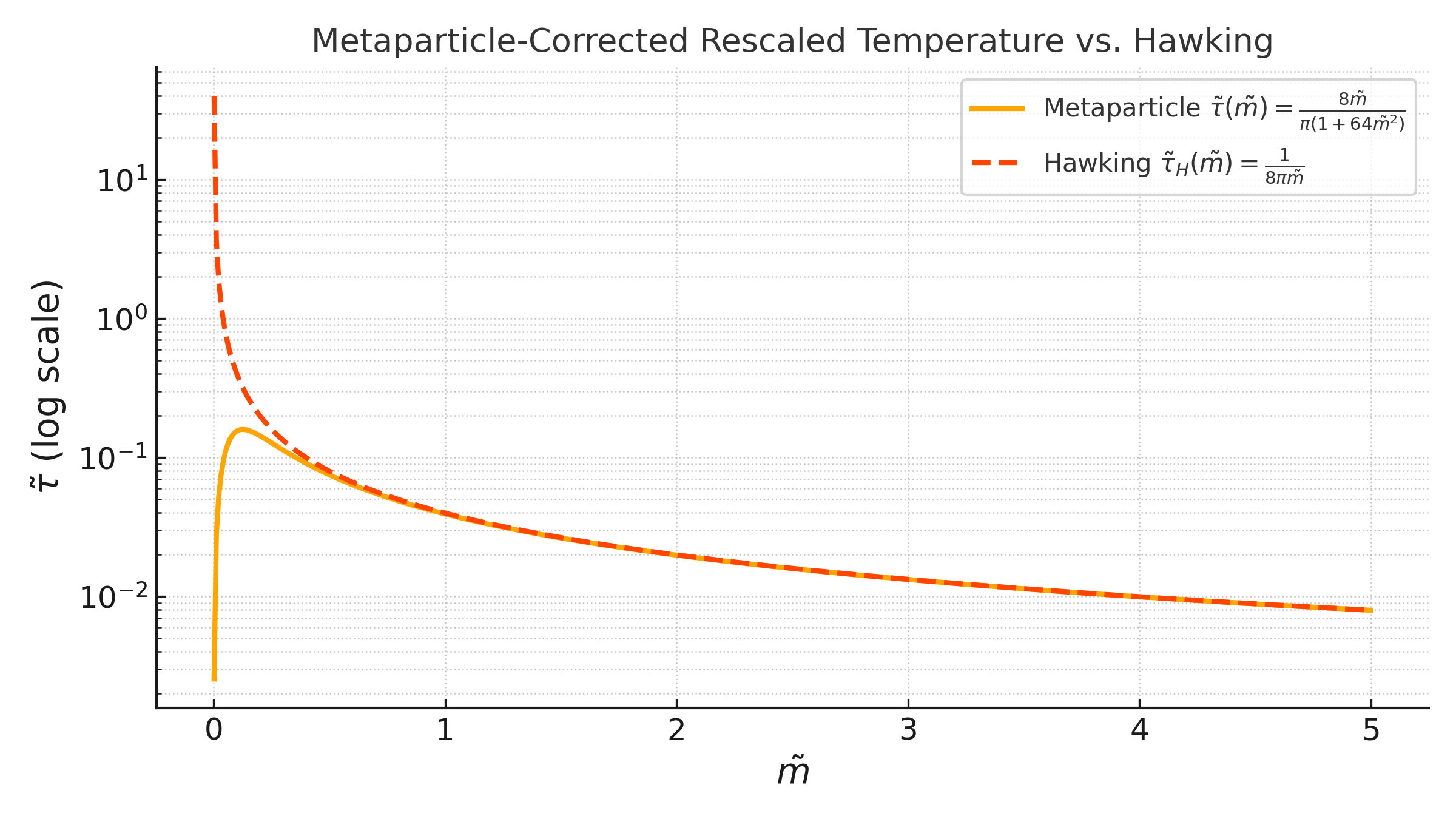}
\includegraphics[width=0.48\textwidth]{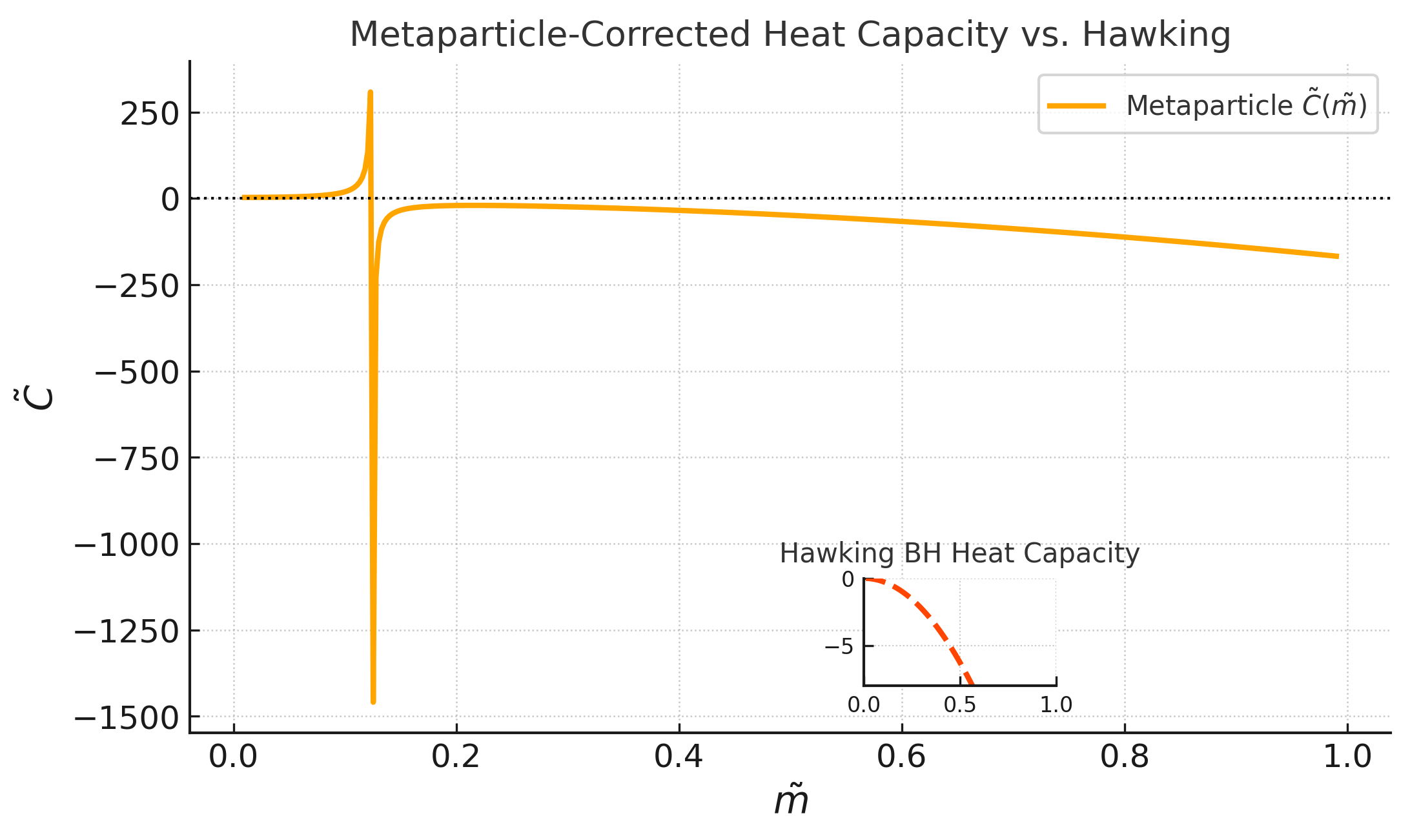}
\caption{Left: The Metaparticle-corrected Hawking (rescaled) temperature $\ttau(\tm)$ showing a finite maximum at $\tau_{\text{max}}=1/2\pi$ followed by a rapid decrease to zero. Right: Comparison between the rescaled metaparticle-corrected heat Capacity (vs rescaled mass $\tm$) with that of Hawking's heat capacity result.}
\label{Fig:Meta-Temp-Heat-Capacity}
\end{figure}
\subsubsection{Emission Rate and Evaporation Time of the Metaparticle-Corrected Black Hole}\label{Sec:Evaporation-Time-1}
In order to get the new picture of the evaporation process of the metaparticle modified black hole (yet without entanglement), we use a Hawking-like mass loss rate with metaparticle-corrected temperature (\ref{Eq:Temp-Meta-No-int}):
\begin{equation}\label{Eq:Emmision-rate-1}
\frac{d\tM}{d\tilde{t}} = -\sigma \cdot A(\tM) \cdot \tT^4(\tM),
\end{equation}
where \( A(\tM) = 16\pi \tM^2 \) is the horizon area, and \( \sigma \) is an effective Stefan-Boltzmann-like constant that includes greybody factors:
\begin{equation}
\sigma = \sigma_{\mathrm{SB}} \cdot \epsilon_{\mathrm{eff}} = \frac{\pi^2}{60} \cdot \epsilon_{\mathrm{eff}}.
\end{equation}


The factor \( \epsilon_{\mathrm{eff}} \) accounts for graybody effects, namely the partial transmission of radiation through the black hole’s curved spacetime potential barrier. These effects suppress the emitted flux relative to a perfect blackbody. Detailed calculations by Don Page \cite{Page:1976df} for black holes radiating Standard Model particles suggest:
\[
\epsilon_{\mathrm{eff}} \approx 0.01,
\]
leading to:
\[
\sigma = \frac{\pi^2}{6000} \approx 1.645 \times 10^{-3} \quad \text{(in Planck units)}.
\]


The evaporation time $\ttau_{\text{evap}}$ is computed by integrating the inverse of the mass loss rate:
\begin{equation}
\ttau_{\text{evap}} = \int_{\tilde{t}_0}^{\tilde{t}_\mathrm{rem}} d\tilde{t}=\int_{\tM_{\mathrm{rem}}}^{\tM_0} \frac{d\tM}{\left| \frac{d\tM}{dt} \right|}.
\end{equation}
where we flipped the integration limits given that the black hole is losing mass $d\tM/d\ttt<0$.
Here we write $\tM_0$ for the initial mass of the black hole and $\tM_{\mathrm{rem}}$ for the resulting remnant mass which can be zero if there is no mechanism preventing the complete evaporation of the black hole. \\
Substituting the expressions for \( A(\tM) \) and \( \tT(\tM) \), we obtain:
\begin{align}
\ttau &= \frac{1}{16\pi \sigma} \int_{\tM_{\mathrm{rem}}}^{\tM_0}
\left(8\pi \tM + \frac{1}{8\tilde{\mu} \tM} \right)^4 \cdot \frac{1}{\tM^2} \, d\tM.
\end{align}


Define:
\[
f(\tM) = \left( 8\pi \tM + \frac{1}{8\tilde{\mu} \tM} \right)^4
= \sum_{n=0}^4 \binom{4}{n} (8\pi \tM)^{4-n} \left( \frac{1}{8\tilde{\mu} \tM} \right)^n
= \sum_{n=0}^4 \binom{4}{n} \frac{(8\pi)^{4-n}}{(8\tilde{\mu})^n} \tM^{4 - 2n}.
\]
So the total evaporation time becomes:
\begin{equation}
\ttau_{\text{evap}} = \frac{1}{16\pi \sigma} \sum_{n=0}^4
\binom{4}{n} \frac{(8\pi)^{4-n}}{(8\tilde{\mu})^n}
\int_{\tM_{\mathrm{rem}}}^{\tM_0} \tM^{2 - 2n} d\tM.
\end{equation}

The integral is elementary and we obtain
\begin{equation}\label{Eq:Tot-Evap-1}
\ttau_{\text{evap}} = \frac{1}{16\pi \sigma} \sum_{n=0}^4
\binom{4}{n} \frac{(8\pi)^{4-n}}{(8\tilde{\mu})^n\,(3-2n)}
\left( \tM_0^{3-2n}-\tM_{\mathrm{rem}}^{3-2n} \right).
\end{equation}

The above expression provides the total evaporation time from an initial mass \( \tM_0 \) down to the remnant mass \( \tM_{\mathrm{rem}} \) for any chosen \( \tilde{\mu} \), assuming that our metaparticle-corrected entropy (\ref{Eq:STot-noEntg}) governs the black hole's thermodynamics.


Compared to the standard Hawking evaporation time with an entropy purely quadratic as a function of mass :
$
S_{\text{Hawking}}(\tM) = 4\pi \tM^2,
$ and thus a temperature 
$
\tT_{\text{Hawking}}(\tM) = \dfrac{1}{8\pi \tM},
$ the total evaporation time 
corresponds to the $n=0$ term in (\ref{Eq:Tot-Evap-1}).
\subsection*{Metaparticle-Corrected Case with \( M_{\mathrm{rem}} \to 0 \)}

If we allow the metaparticle-corrected black hole to evaporate fully to zero mass (i.e., \( \tM_{\mathrm{rem}} \to 0 \)), the evaporation time (\ref{Eq:Tot-Evap-1}) diverges for the \( n = 2 \) term, where \( 2 - 2n = -2 \), and more strongly for \( n = 3, 4 \), making:
\begin{equation}
\ttau_{\text{meta}} \to \infty.
\end{equation}

\begin{figure}[h!]
\centering
\includegraphics[width=0.9\textwidth]{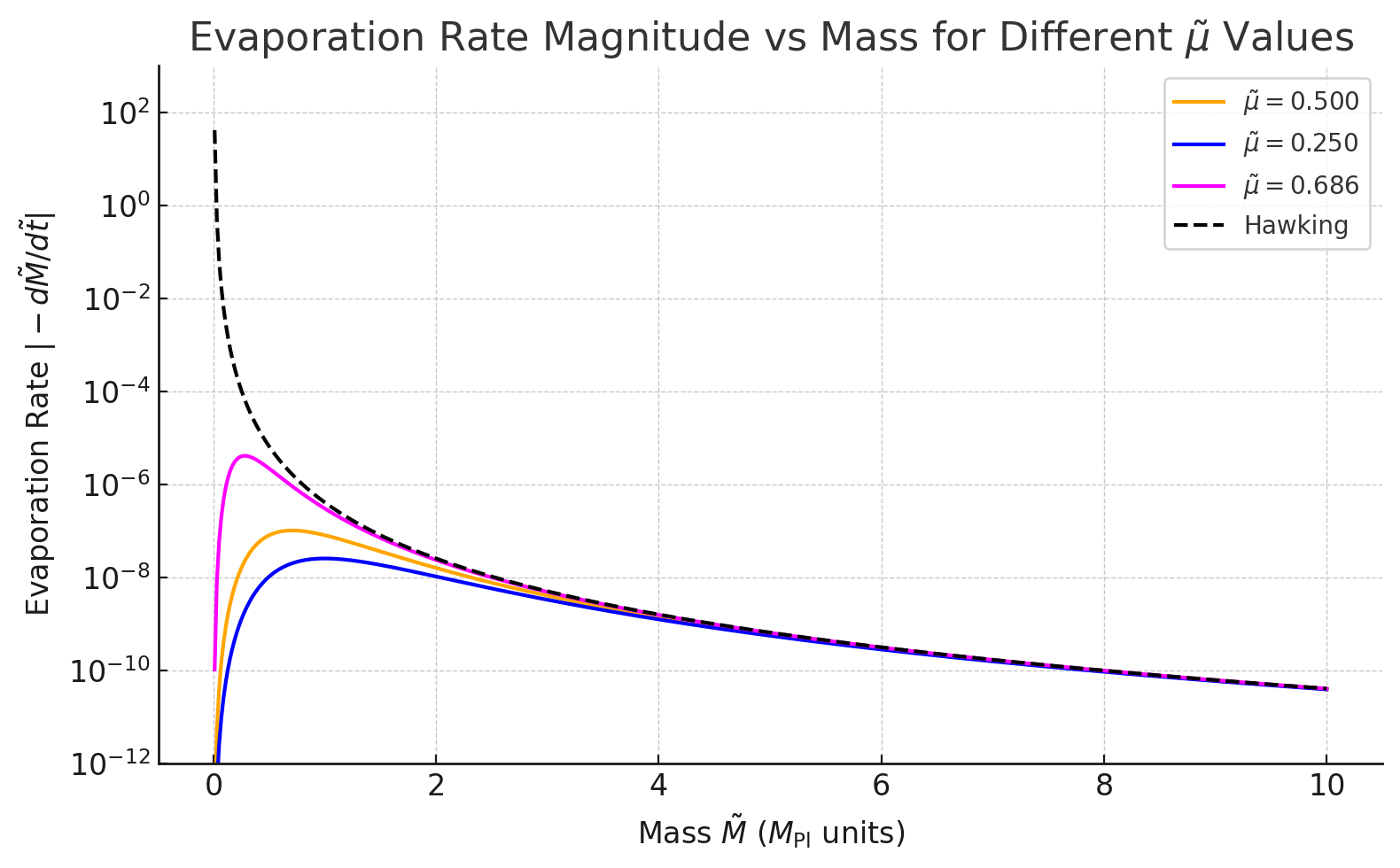}
\caption{Evaporation rates \( -\frac{d\tilde{M}}{d\tilde{t}} \) for various values of \( \tilde{\mu} \), compared to the Hawking case (dashed line). Bottom: total evaporation time for each model, assuming \( M_0 = 10 M_P\).}
\label{Fig:Evap-Rate-NonEnt}
\end{figure}

Thus, in contrast with the Hawking case (which yields a finite lifetime even when \( \tM_{\mathrm{rem}} = 0 \)), the metaparticle-corrected black hole cannot evaporate completely in finite time. This is due to the logarithmic correction in the entropy, which causes the temperature to vanish faster at small \( \tM \), effectively halting evaporation and suggesting the existence of a remnant or frozen final state.
This can also be seen from figure \ref{Fig:Evap-Rate-NonEnt}. As the black hole evaporates and reaches smaller and smaller masses, the rate of evaporation goes to zero suggesting equally a shut down of the Hawking radiation.


So, while Hawking's computation predicts complete evaporation in finite time, metaparticle-stringy corrections imply that black holes asymptotically approach a remnant mass, and the evaporation time diverges as \( M_{\mathrm{rem}} \to 0 \). This supports the idea that quantum gravity effects could prevent full black hole evaporation and resolve the information paradox through remnant scenarios.

\par However, although from the analysis in this section, we get an improved physical picture of the possible end-life of an evaporating black hole, still allowing the remnant mass to go to zero is incompatible with the existence of a minimal length supported by many approaches to quantum gravity \cite{Mead1964,Garay1995,KempfManganoMann1995,
GrossMende1988,AmatiCiafaloniVeneziano1989,RovelliSmolin1995,
DoplicherFredenhagenRoberts1995}. GUP analysis as reviwed in \cite{Hossenfelder2013} and the Extended GUP we derive in \cite{Chouha-PR-cosm:2026-c} and its analysis point also to such a minimal length.
Furthermore,  metastring theory, of which metaparticle are the zero modes, has an intrinsic minimal length built into it and supported by our analysis in section \ref{sec:interpretation} (c.f. equation (\ref{Eq:Min_Len_Ent})). 
\par A closer look at the behavior of the entropy as $\tM\to0$ in figure \ref{fig:metaparticle_entropy} shows that the entropy negative which is nonsensical. More precisely, the entropy becomes negative once the mass of the black hole dips below the critical mass \ref{Eq:Critical-mass-no-Ent}. This is supposed to be where the phase transition occurs and where the heat capacity becomes positive stabilizing the system. We conclude then that this stability can not be reached. Adding to that, as $\tM\to0$ the heat capacity \ref{Eq:Heat_Cap} remains finite, signaling an incompatibility with the temperature being zero.
and thus the above scenario cannot be the end of the story. 
\par
The crucial physical ingredient that is still missing is a treatment of the metaparticle and its dual as a single quantum object, rather than as two independent degrees of freedom. 
In other words, the analysis so far neglects the intrinsic correlations---or, more precisely, the entanglement---between the metaparticle sector and its dual sector. 
In the following, we incorporate this entanglement explicitly. 
As we shall see, doing so leads to a consistent and physically coherent picture.

\subsection{Completing the Picture: Bekenstein's Argument in Light of Metaparticles With Entanglement}\label{subsec:Bek-Meta-Ent}
Recall that our previous cosmological analysis of the dispersion relations and hence the equation of state of both the metaparticle (\ref{Eq:w-param-metapart}) and the dual-metaparticle (\ref{Eq:w-param-dual-metapart}) allowed us to identify the metaparticle as behaving as momentum modes (in the regular stringy interpretation) and the dual-metaparticle as winding modes in the high energy regime. Thus the entropy of the metaparticle $S^{+}(A)$ is associated with the geometric phase of the black hole evolution while the entropy of the dual-metaparticle $S^{-}(A)$ with the non-geometric phase.
We previously found the entropy of the two branches:\\
(\textbf{The dual non-geometric-branch})
\begin{equation}\label{eq:Sm}
S^{-}(A) = \frac{\ln 2}{2\pi\,\Gamma} \left[ A + \frac{\pi}{2\tilde{\mu}} \ln\left( 1 + \sqrt{1 + \frac{4\tilde{\mu} A}{\pi}} \right) + \frac{\pi}{2\tilde{\mu}} \sqrt{1 + \frac{4\tilde{\mu} A}{\pi}} \right] \text{and}
\end{equation}
(\textbf{The geometric branch})\footnote{Recall that the two entropy branches $S^{+}$ and $S^{-}$ correspond respectively to two complementary sectors of the underlying phase--space degrees of freedom. The branch $S^{+}$ is termed geometric because it supports an emergent spacetime interpretation in the infrared, while $S^{-}$ captures the dual (winding-like) sector that becomes dominant in the non-geometric regime.
}\\
\begin{equation}\label{eq:Sp}
 S^{+}(A) = \frac{\ln 2}{2\pi\,\Gamma} \left[ A + \frac{\pi}{2\tilde{\mu}} \ln\left( \frac{4\tilde{\mu} A}{\pi} \right) - \frac{\pi}{2\tilde{\mu}} \ln\left( 1 + \sqrt{1 + \frac{4\tilde{\mu} A}{\pi}} \right) - \frac{\pi}{2\tilde{\mu}} \sqrt{1 + \frac{4\tilde{\mu} A}{\pi}} \right].
\end{equation}
where \( A \) is once again the horizon area in Planck units and \( \Gamma \) a new normalization factor that we will fix by requiring as before the large area behavior of the entropy to coincide with the well known Hawking result.
\subsubsection{Pseudo-Entanglement Total Entropy}
We now consider the true nature of the metaparticle. What we call a metaparticle is really a single quantum object made of a particle entangled with its dual as represented pictorially by the quantum state (\ref{Eq:Ket-meta}), and not as two separate particles, which exist independently of each other. 

To do so, we propose the following new expression for the total entropy, incorporating a cross-term interpreted as the entanglement/correlation entropy between these two sectors (the geometric and non-geometric sectors). 

We define the \emph{pseudo-entanglement total entropy} of the system as:
\begin{align}
S_{\text{Tot}}^{\text{Ent}}(A) = \left( \sqrt{S^+(A)} + \sqrt{S^-(A)} \right)^2 = S^+(A) + S^-(A) + 2\sqrt{S^+(A) S^-(A)}. \label{eq:Stot}
\end{align}

The cross-term,
\begin{align}
S_{\text{mix}}(A) \equiv 2\sqrt{S^+(A) S^-(A)},\label{Eq:SMix}
\end{align}
is proposed as a measure of the \emph{quantum entanglement/correlations} between the metaparticle and its dual. This structure mirrors quantum interference terms and aligns with intuition from thermofield double states and holographic entanglement entropy
\cite{Maldacena:2001kr,VanRaamsdonk:2010pw} as well as captures the correlations between the metaparticle and its dual mimicking the construction in \cite{Volovik:2021upi}.
\par 
The form chosen for the total entropy is not intended to represent a Von Neumann entanglement entropy derived from an explicit density matrix, which would require a full microscopic Hilbert-space construction beyond the scope of the present work. Rather, it should be viewed as the minimal thermodynamically consistent completion of the entropy once the metaparticle is treated as a single quantum object composed of correlated geometric and dual degrees of freedom.

The proposed structure is uniquely fixed by a small set of physically motivated requirements: (i) symmetry under interchange of the geometric and dual sectors, reflecting the underlying duality of the metaparticle; (ii) reduction to the additive entropy $S^+ + S^-$ in the absence of correlations; (iii) positivity and extensivity of the total entropy in the large-area (infrared) limit, ensuring recovery of the Bekenstein–Hawking area law; and (iv) sensitivity to correlations near the ultraviolet regime, where one of the entropy branches vanishes and entanglement effects become dominant. The square-root structure ensures that the mixed term becomes leading precisely when the independent-sector description breaks down, while remaining subleading at large horizon area.
\par Among the four criteria listed above, the first three constrain the individual
geometric and dual entropy contributions and ensure the correct semiclassical and
thermodynamic limits in each sector separately. They do not, however, determine how
the two sectors are coupled. The fourth criterion fixes this coupling and uniquely
selects the cross-term appearing in Eq.~(\ref{eq:Stot}). Once a nontrivial coupling is required,
duality symmetry severely restricts its functional form: the cross-term must be
symmetric under interchange of $S^{+}$ and $S^{-}$, scale homogeneously with entropy,
and vanish smoothly when either sector is suppressed. The square-root structure
$\sqrt{S^{+}S^{-}}$ is the minimal invariant satisfying these requirements. Moreover,
this choice is singled out by stability and consistency: it correlates fluctuations
between the two sectors without introducing runaway behavior, preserves the
concavity of the total entropy, and is the minimal coupling that guarantees the
reality and global well-definedness of the physical entropy over the entire domain
of interest. In this sense, the fourth criterion does not introduce an ad hoc
structure, but fixes the unique minimal cross-term compatible with duality symmetry,
stability, and entropy reality, making explicit a coupling that is already implicit
in doubled and modular formulations of spacetime and in non-factorizing Hilbert
spaces \cite{Polchinski:1998rq,Hull:2009mi,FreidelLeighMinic2015}.

From this perspective, the pseudo-entanglement entropy should be understood as an effective, coarse-grained measure of correlations between the two sectors of the metaparticle phase space. Its role is not to introduce new microscopic degrees of freedom, but to encode the fact that geometric and dual excitations cannot be treated as effectively uncorrelated once the black hole approaches the regime where a single spacetime polarization ceases to be globally valid.


\par Demanding $S_{\text{Tot}}^{\text{Ent}}(A)\sim A/4$ for $A\to\infty$ fixes
\[
\Gamma=\frac{8\ln2}{\pi},
\]
and thus the total expression reads
\begin{equation}
\!S_{\text{Tot}}^{\text{Ent}}(A)=\frac{1}{16}\!\Bigl[
  2A+\alpha\ln\!\Bigl(\frac{4\tilde{\mu}A}{\pi}\Bigr)
  +2\sqrt{\bigl[A+\alpha(\ln(1+\chi)+\chi)\bigr]
           \bigl[A+\alpha(\ln(\tfrac{4\tilde{\mu}A}{\pi})
                     -\ln(1+\chi)-\chi)\bigr]}
\Bigr]
\label{eq:S_tot_A}
\end{equation}
where 
\[
\alpha \equiv \frac{\pi}{2\tilde{\mu}}, \qquad
\chi \equiv \sqrt{1+\frac{4\tilde{\mu}A}{\pi}} .
\]
As a consistency check, for $A \gg \pi/(4\tilde{\mu})$ one has $\chi \simeq 2\sqrt{\tilde{\mu}A/\pi}$ and
\[
S_{\text{Tot}}^{\text{Ent}}(A)
  =\frac{A}{4}
  +\frac{\pi}{16\tilde{\mu}}
   \Bigl[\ln\!\Bigl(\tfrac{4\tilde{\mu}A}{\pi}\Bigr)-1\Bigr]
  +\mathcal O(A^{-1/2}).
\]
which is our previous result (\ref{Eq:STot-noEntg}) up to a constant.
\par We next introduce dimensionless physical quantities to simply the expressions and analysis of the thermodynamical functions. Thus, we let:
\[
\tilde a\equiv\frac{4\tilde{\mu}A}{\pi},\qquad
\chi(\tilde a)\equiv\sqrt{1+\tilde a},\qquad
\mathcal F(\tilde a)\equiv\ln(1+\chi)+\chi.
\label{Eq:rescaled-a}\]

\begin{equation}
S_{\text{Tot}}^{\text{Ent}}(\tilde a)=
\frac{\pi}{32\tilde{\mu}}\Bigl[
  \tilde a+\ln\tilde a
  +2\sqrt{\Bigl(\tfrac{\tilde a}{2}+\mathcal F\Bigr)
           \Bigl(\tfrac{\tilde a}{2}+\ln\tilde a-\mathcal F\Bigr)}
\Bigr].
\label{eq:S_tot_a}
\end{equation}
\begin{figure}[h!]
  \centering
  \includegraphics[width=0.8\textwidth]{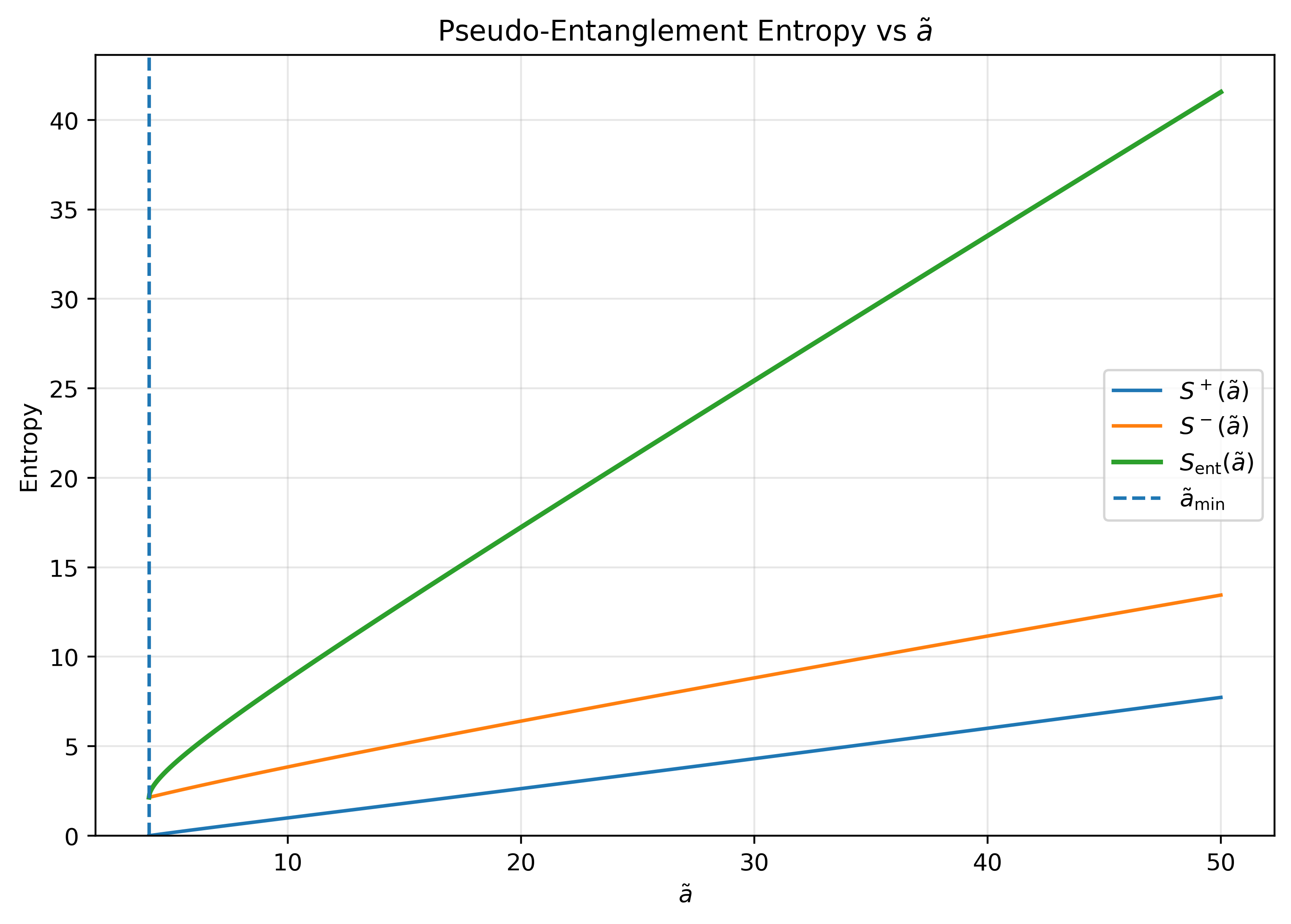}
  \caption{Pseudo-entanglement entropy $S_{\mathrm{ent}}(\tilde a)$ (green) 
  compared with the individual branches $S^+(\tilde a)$ and $S^-(\tilde a)$. 
  The vertical dashed line marks the minimum allowed area 
  $\tilde a_{\min}$, below which the entangled entropy becomes non-real.}
  \label{fig:pseudo-entanglement-entropy}
\end{figure}

\begin{figure}[h!]
  \centering
  \includegraphics[width=0.85\textwidth]{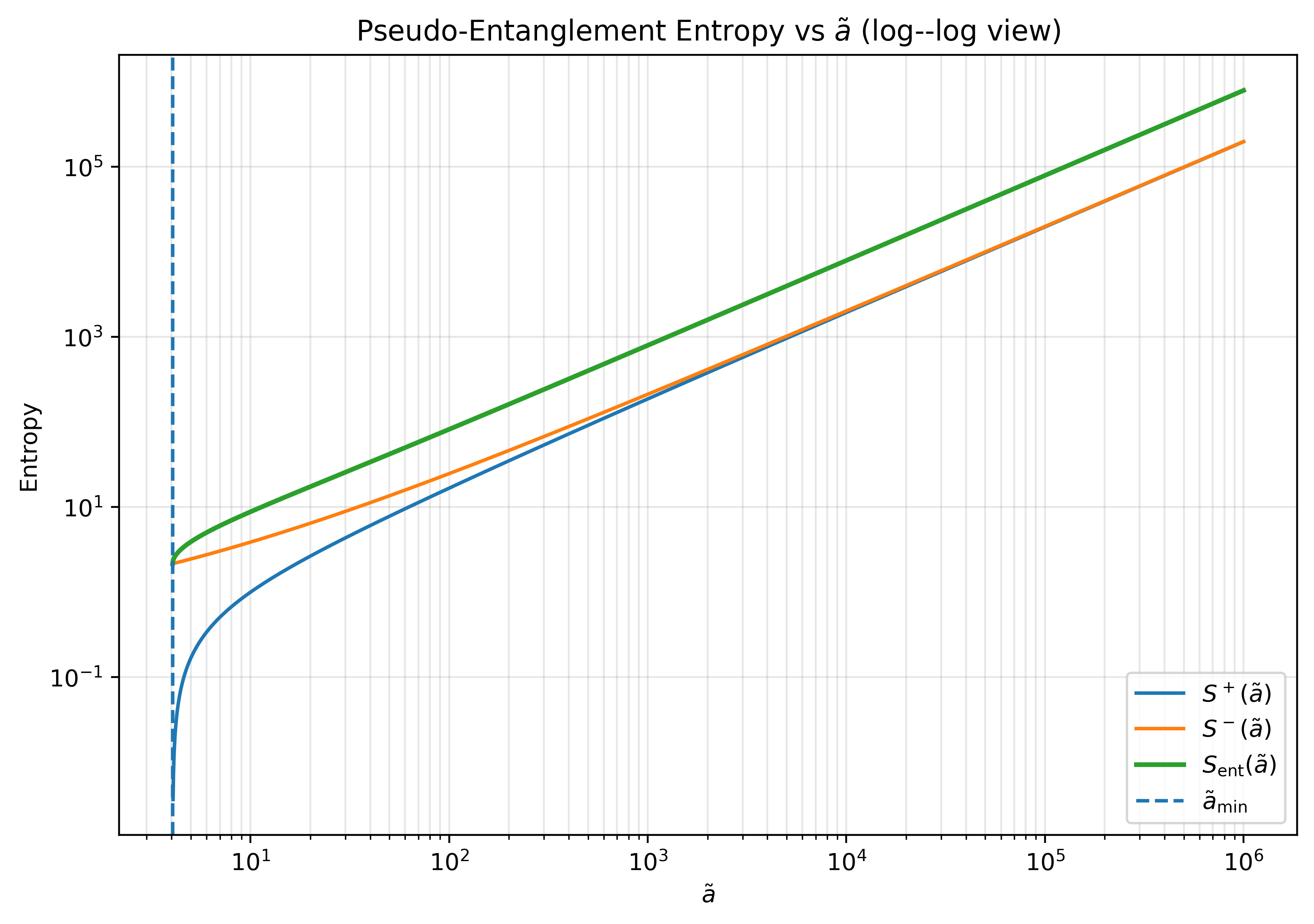}
  \caption{Log--log plot of the pseudo-entanglement entropy $S_{\mathrm{ent}}(\tilde a)$ (green)
  together with the individual branches $S^+(\tilde a)$ and $S^-(\tilde a)$, shown up to
  very large $\tilde a$. The vertical dashed line marks the dynamically generated minimum
  area $\tilde a_{\min}$, below which the entangled entropy becomes non-real. The log--log
  representation makes visible the suppressed growth of the $S^+$ branch and shows that,
  at sufficiently large $\tilde a$, all three entropies approach parallel linear behavior,
  signaling the recovery of the classical Bekenstein--Hawking scaling.}
  \label{fig:pseudo-entanglement-large-a-loglog}
\end{figure}
\subsubsection{A Minimal Area $\tA_\text{min}$ (and Length) and its Interpretation} 
\label{sec:interpretation}
We also define
\[
 g(\tilde{a}) := \frac{\tilde{a}}{2} + \ln\tilde{a} - \mathcal{F}(\tilde{a}).
\]
The square root in~\eqref{eq:S_tot_a} is real only if $g(\tilde{a}) \ge 0$. The unique positive root of $g$ is
\begin{equation}\label{Eq:a_min}
\tilde{a}_\text{min}=W(e^2)+2\sqrt{W(e^2)} \approx 4.0523.
\end{equation}
where $W(x)$ refers to the Lambert function\footnote{The Lambert $W$ function is defined implicitly by $W(x)e^{W(x)}=x$ \cite{Corless1996LambertW}.}.
Hence the exact entropy is well--defined for
\begin{equation}\label{Eq:Minimal_Area_Entg}
\boxed{\tilde a\;\ge\;\tilde a_{\text{min}}\;\approx\;4.053}
\quad\Longleftrightarrow\quad
\boxed{\tA\;\ge\;\tA_{\text{min}}= \frac{\pi\,\tilde a_{\min}}{4\tilde{\mu}} }.
\end{equation}

Recall that the two entropy branches $S^{+}$ and $S^{-}$ were defined, corresponding respectively to ``geometric'' and ``dual'' sectors of the underlying phase--space degrees of freedom. 

Below $\tA_{\text{min}}$ the branch $S^{+}$ becomes negative, rendering $\sqrt{S^{+}}$ imaginary and invalidating the semiclassical construction. Hence $\tA_{\text{min}}$ is the \emph{smallest} geometric surface capable of supporting \emph{both} families of metaparticle microstates, i.e. both geometric and dual families of microstates respectively; below this scale, geometric degrees of freedom cease to exist as physical states and only dual (non-geometric) excitations remain meaningful.\\[4mm]

Thus, the minimal area \ref{Eq:a_min} gives rise to a minimal length scale
\begin{equation}\label{Eq:Min_Len_Ent}
    \boxed{\tl_\text{min}\gtrsim \frac{1}{2\sqrt{\tmu}}}
\end{equation}
The existence of a minimal length is crutial in resolving the physical inconsistencies mentioned at the end of last section \ref{Sec:Evaporation-Time-1}.
\par To complete the thermodynamical history of the metaparticle-corrected black hole we need both the temperature and the heat capacity which we compute next.

\subsubsection{Temperature and Heat Capacity}
We work in the rescaled variables introduced in (\ref{Eq:rescaled-vars}) and assuming that the surface area of the event horizon to still be given by $4\pi  r_s^2$ with the regular Schwarzschild radius $r_s=2M$ we have:
\begin{equation}\label{Eq:rescaled_a-m-dep}
\boxed{\tilde a = 64\,\tilde m^{\,2}}
\end{equation}


The exact total entropy (\ref{eq:S_tot_a}) in in terms of the rescaled mass $\tm$ now reads:

\[
\boxed{%
S_{\text{Tot}}^{\text{Ent}}(\tilde m)
 =\frac{\pi}{32\tilde\mu}
  \Bigl[
     64\tilde m^{2}
     + \ln\!\bigl(64\tilde m^{2}\bigr)
     +2\sqrt{\Bigl(32\tilde m^{2}+\mathcal F(\tilde m)\Bigr)
             \Bigl(32\tilde m^{2}+\ln(64\tilde m^{2})-\mathcal F(\tilde m)\Bigr)}
  \Bigr]},
\]
where
\[
\chi(\tilde m)=\sqrt{1+64\tilde m^{2}},
\qquad
\mathcal F(\tilde m)=\ln\!\bigl(1+\chi(\tilde m)\bigr)+\chi(\tilde m).
\]

This expression contains only the duality-parameters $\tilde\mu$ and the mass‐like variable $\tilde m$.


\par We proceed similarly to our previous analysis by defining a temperature independent of $\tmu$.\\
Define
\begin{equation}\label{Eq:Def-of-dimensionless-temp}
\boxed{\tilde\tau(\tilde m)
       \;=\;
       \frac{\tilde T(\tilde m)}{\sqrt{\tilde\mu}}
       \;=\;
       \frac{T(\tilde m)}{M_P\sqrt{\tilde\mu}} }.
\end{equation}

We also define the following functions and their derivatives to simplify the expression of the temperature: 
\[
B(\tilde m)         = \ln\!\bigl(64\tilde m^{2}\bigr),\quad
C(\tilde m)         = 32\,\tilde m^{2},\quad
D_1(\tilde m)       = C+\mathcal F,\quad
D_2(\tilde m)       = C+B-\mathcal F,\quad
Z(\tilde m)         = \sqrt{D_1D_2}\,.
\]
\[
\chi'          = \frac{64\tilde m}{\chi},\quad
(\mathcal F)'  = \frac{32\tilde m}{\chi}\,
                 \frac{2+\chi}{1+\chi},\quad
B'             = \frac{2}{\tilde m},\quad
C'             = 64\tilde m,
\]
\[
\begin{aligned}
D_1' &= 64\tilde m+\frac{32\tilde m}{\chi}\,
                      \frac{2+\chi}{1+\chi},\\
D_2' &= 64\tilde m+\frac{2}{\tilde m}-\frac{32\tilde m}{\chi}\,
                      \frac{2+\chi}{1+\chi},\\
Z'   &= \frac{D_2D_1'+D_1D_2'}{2Z}.
\end{aligned}
\]

Thus 
\[
\frac{dS}{d\tilde m}
  =\frac{\pi}{32\tilde\mu}
   \Bigl[\,128\tilde m+\frac{2}{\tilde m}+2Z'(\tilde m)\Bigr].
\]

and finally the rescaled temperature reads:
\begin{equation}
\boxed{%
\tilde\tau(\tilde m)
  =\frac{32}{\pi}\;
   \Bigl[\,128\tilde m+\dfrac{2}{\tilde m}+2Z'(\tilde m)\Bigr]^{-1}}. \label{Eq:Temp_Ent}
\end{equation}
For $\tilde m\gg1$ one has $Z'(\tilde m)=\mathcal O(\tilde m)$, so
\[
\tilde\tau(\tilde m)\;\xrightarrow{\;\tilde m\to\infty\;}
\frac{1}{8\pi\,\tilde m}\,
\Bigl[1+\mathcal O\!\bigl(\tfrac{\ln\tilde m}{\tilde m^{2}}\bigr)\Bigr],
\]
i.e.\ the usual Hawking behaviour $T=1/(8\pi M)$ once
$M=\tilde m/\sqrt{\tilde\mu}$ is restored.


\subsubsection{Behavior of the Temperature}
\begin{figure}[h!]
\centering

\includegraphics[width=0.7\textwidth]{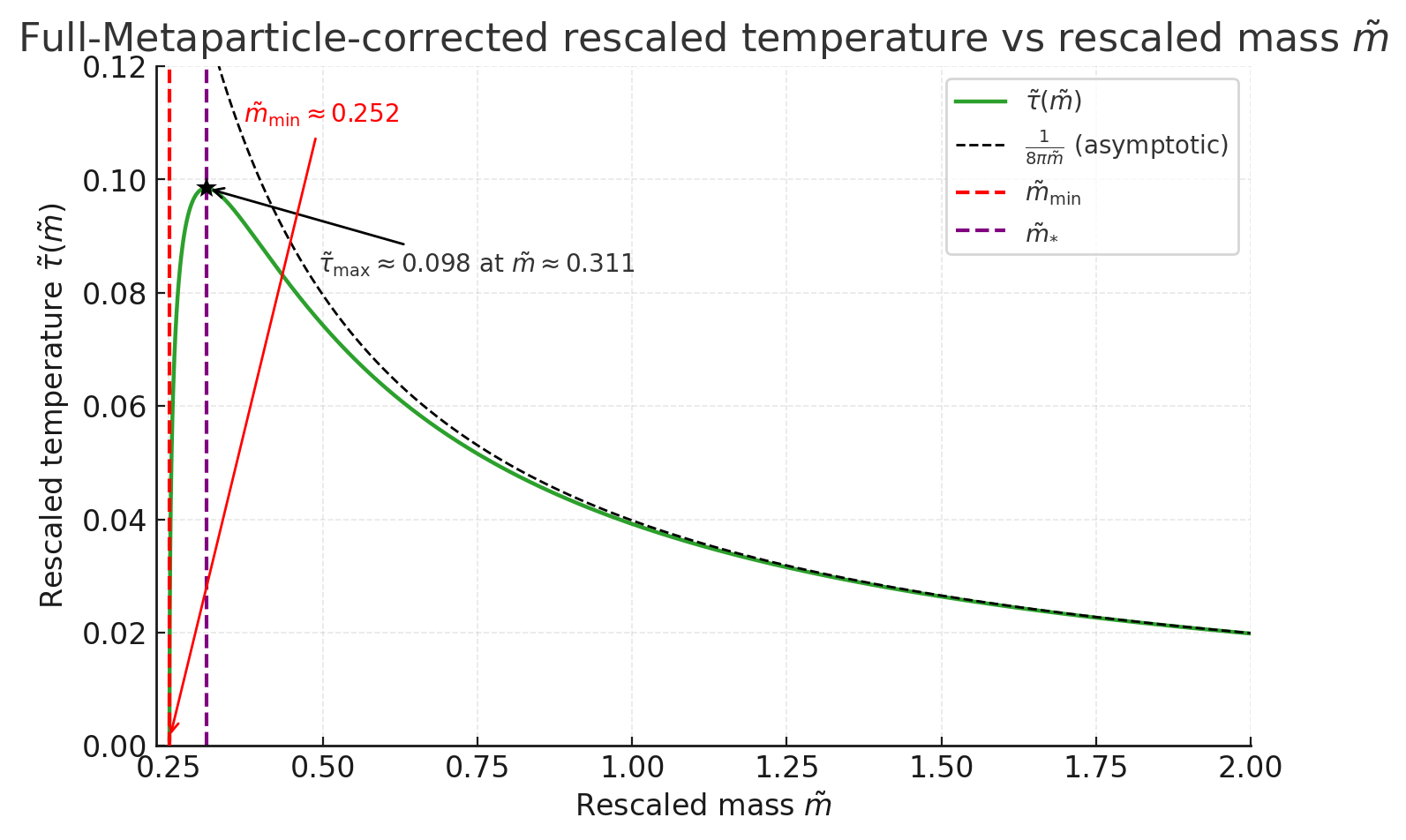}
\caption{Full metaparticle-corrected rescaled temperature \( \tilde{\tau}(\tilde{m}) \) as a function of \( \tilde{m} \).}
\label{Fig:Temp_Ent}
\end{figure}

As shown in the plot of Fig.~\ref{Fig:Temp_Ent} the rescaled temperature \( \tilde{\tau}(\tilde{m}) \) (\ref{Eq:Temp_Ent}) rises as the black hole mass decreases, reaching a maximum $\ttau_\text{max}(\tm_*)\approx0.098$ around \( \tilde{m_*} \approx 0.311 \), then decreases towards zero as we reach \(\tilde{m}_\text{min} = \sqrt{\tilde{a}_\text{min}/64} \approx 0.252 \), where \( \tilde{a}_{\min} \approx 4.053 \) is the minimal area (\ref{Eq:Minimal_Area_Entg}) allowed by the pseudo-entangled entropy formula.

More precisely, as 
\(\tilde m \to \tilde m_{\min}^+\), $\tilde{\tau}(\tilde m) \;\longrightarrow\; 0^+$,
with a leading asymptotic given by\footnote{Let $\delta=\tilde m-\tilde m_{\min}\downarrow0^+$. Since $\hat S^+(\tilde m_{\min})=0$
and $\hat S^+$ is differentiable with $s_1\equiv d\hat S^+/d\tilde m|_{\tilde m_{\min}}\neq0$,
Taylor expansion gives
\[
\hat S^+(\tilde m)=s_1\delta+o(\delta).
\]
Hence the mixed term dominates near the edge and
\[
\hat S_{\rm Tot}=S^-_0+2\sqrt{S^-_0\,\hat S^+}+o(\delta^{1/2})
= S^-_0 + 2\sqrt{S^-_0\,s_1}\,\delta^{1/2}+o(\delta^{1/2}).
\]
Therefore
\[
\frac{d\hat S_{\rm Tot}}{d\tilde m}=\sqrt{S^-_0\,s_1}\,\delta^{-1/2}+o(\delta^{-1/2}),
\]
and with $\tilde\tau=(1/\pi)(d\hat S_{\rm Tot}/d\tilde m)^{-1}$ we obtain
\[
\tilde\tau(\tilde m)=\frac{1}{\pi\sqrt{s_1\,S^-_0}}\,\delta^{1/2}[1+o(1)]
\propto\sqrt{\tilde m-\tilde m_{\min}}.
\]
} 

\begin{equation}\label{Eq:Asympt-Tem-Entg}
\tilde{\tau}(\tilde m)\;\approx\;1.139\,\sqrt{\tilde m-\tilde m_{\min}}
\qquad \text{as }\tilde m \downarrow \tilde m_{\min} .
\end{equation}

\par Comparing our results for the full temperature of the evaporating black hole taking into account the quantum correlations between the metaparticle and its dual versus without entanglement we see that although the two respective plots Fig.~\ref{Fig:Temp_Ent} and Fig.~\ref{Fig:Meta-Temp-Heat-Capacity} are qualitatively the same, the major difference is that the temperature in the latter case goes to zero as the black hole is allowed to asymptotically evaporate completely while the minimal allowed area coming from the quantum entanglement between the metaparticle and its dual prevents such a total evaporation and we end up with cold zero-temperature remnant. To complete the picture  we need the input of the heat capacity which will tell us what the black hole is thermally doing as it reaches its final stages.

\subsubsection{Heat Capacity and Phase Transition}
\begin{figure}[ht]
\centering
\includegraphics[width=0.7\textwidth]{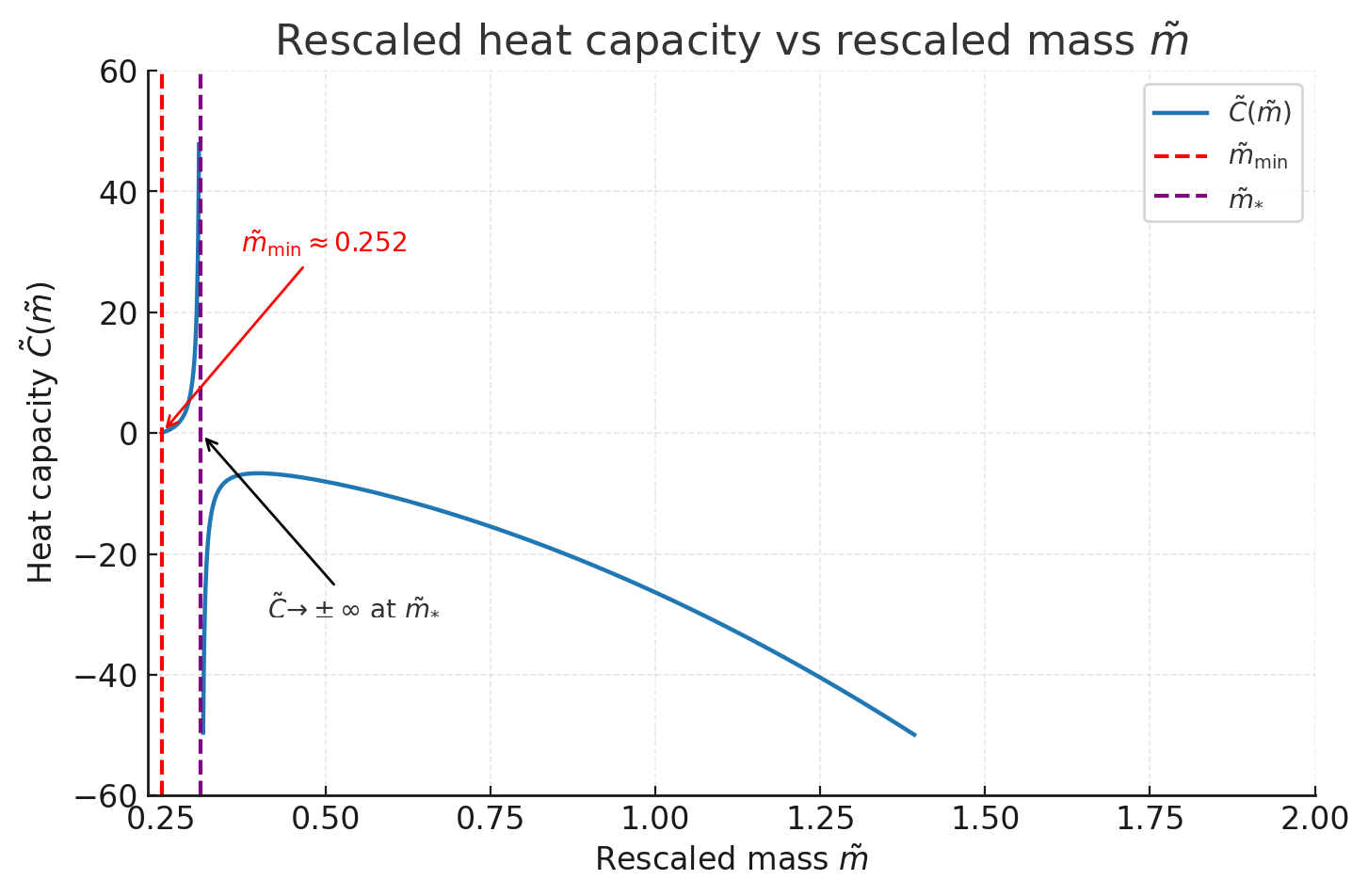}
\caption{Dimensionless heat capacity \( \tilde{C}(\tilde{m}) \) as a function of \( \tilde{m} \).
    The red dashed line marks the minimal mass $\tilde m_{\min}\!\approx 0.252$, 
    where the heat capacity vanishes. 
    The purple dashed line marks $\tilde m_{*}\!\approx 0.311$, 
    where $\tilde\tau$ attains its maximum and $\tilde{C}$ diverges. 
    The plot uses the same scale and units as that of the temperature Fig.~\ref{Fig:Temp_Ent} for direct comparison.}
\label{fig:Heat-Capacity-Entng}
\end{figure}

In what follows, we do not explicitly give the full formula for the heat capacity, since it is extremely messy and is not insightful, but we plot it in Fig.~\ref{fig:Heat-Capacity-Entng} and discuss its behavior.

The heat capacity, defined as:
\[ \tilde{C}(\tilde{m}) = \left( \frac{d\tilde{\tau}}{d\tilde{m}} \right)^{-1}, \] behaves as follows:
\begin{itemize}
    \item Near the threshold $\tilde m_{\min}$: since 
    \(
    \tilde\tau(\tilde m) \sim  \sqrt{\tilde m - \tilde m_{\min}}\), one finds
    \[
        \tilde C(\tilde m) \sim 2\sqrt{\tilde m - \tilde m_{\min}} \to 0^+
        \quad (\tilde m \downarrow \tilde m_{\min}).
    \]

    \item At the peak $\tilde m = \tilde m_*$: 
    \(
    \frac{d\tilde\tau}{d\tilde m} = 0 \implies \tilde C \to \pm \infty
    \),
    with a sign flip across the maximum.

    \item For large $\tilde m$: since 
    \(
    \tilde\tau(\tilde m) \sim 1/(8\pi \tilde m)
    \),
    the heat capacity behaves as
    \[
        \tilde C(\tilde m) \sim -8\pi \tilde m^2 ,
        \qquad \tilde m \to \infty.
    \]
\end{itemize}
undergoes a divergence at the temperature maximum. This indicates a second-order phase transition. That is, for \( \tilde{m} > \tilde{m}_*\!\approx 0.311 \), the heat capacity is negative, reflecting the usual black hole instability. For \( \tilde{m} < \tilde{m}_*\), it becomes positive, indicating a stabilizing regime as the black hole enters the non-geometric phase and reaches zero at the minimal allowed mass $\tm_\text{min}\approx0.252$.

At the critical or threshold mass \( \tilde{m}_* \approx0.311 \) (or \(\tM_*=\frac{0.311}{\sqrt{\tmu}})\), the metaparticle-corrected black hole temperature reaches a maximum value:
\[
\tilde{T}_{\text{max}} = 0.098\sqrt{\tmu}\,,
\]
and the corresponding heat capacity diverges at this point:
\[
\lim_{\tilde{m} \to \tilde{m}_c^\pm} \tilde{C}(\tilde{m}_*) \to \pm\infty.
\]
This divergence is a hallmark of a \emph{second-order (continuous) phase transition} in black hole thermodynamics. It marks the critical point at which the black hole transitions from an unstable evaporative phase (with negative heat capacity) to a stable, non-geometric-cooled phase (with positive heat capacity). Unlike first-order transitions, there is no discontinuity in entropy or mass, but the diverging heat capacity indicates critical behavior and possible large thermal fluctuations near \( \tT_{\text{max}} \).

\medskip
While the divergence of the heat capacity at \( \tilde{m}_*\) is characteristic of a second-order phase transition and signals possible large thermal fluctuations, it does not imply a sudden burst of radiation or a breakdown of the black hole. Instead, this divergence arises because the system becomes thermodynamically critical: the temperature reaches an extremum and the derivative \( dT/dM \to 0 \), so the heat capacity \( C = (dT/dM)^{-1} \to \infty \). Physically, this means that the black hole becomes thermally "soft" at \( T_{\text{max}} \), with small changes in mass leading to minimal changes in temperature. Hawking radiation is still present at this point since the temperature remains finite, but the emission rate begins to decrease beyond \( \tilde{m}_*\) as the black hole enters the dual, "quantum"-cooled regime. Thus, the divergent heat capacity reflects a critical instability, not energetic outflow — a thermal signature of the onset of the dual phase transition.
\par It is important to stress that while in the usual Schwarzschild black hole evaporation scenario one expects a significant back-reaction due to the Hawking-radiation, this does not occur in our metarparicle-scenario provided that the duality parameter $\tmu$ is constrained. The back-reaction is significant when the mass scale drops below the thermal energy scale, which could be a problem close to the maximal temperature. However, in our case, in that regime  
\begin{equation}\label{Eq:back-reaction}
\frac{\tT_{\text{max}}}{\tM_{*}}\approx 0.32\, \tmu
    \end{equation}
and the back-reaction remains small provided the duality parameter satisfies the following bound
\begin{equation}\label{Eq:bound-mu}
    \sqrt{\mu}\lesssim 1.8\, E_P
\end{equation}
In Appendix \ref{App:Adiabaticity}, we further discuss the validity of our adiabatic treatment i.e. that the background geometry evolves slowly compared to the dynamical time scales.
\section{Discussion and Comparisons}\label{Sec:Disc-Comparison}

In this section, we synthesize the physical picture suggested by the thermodynamical analysis presented in this work. 
We carefully analyze the implications of each physical conclusion our study suggests and compare them with non-singular remnants arising in mimetic gravity \cite{Chamseddine:2019pux}. 
The motivation for this comparison is that both frameworks lead to remarkably similar late-time thermodynamic behavior for evaporating black holes, including the existence of a maximal temperature~(\ref{Eq:Asympt-Tem-Entg}), a change in the sign of the heat capacity, and a finite critical mass at which geometric Hawking evaporation halts. 
Despite these similarities, the microscopic origin and physical interpretation of the resulting remnants are fundamentally different.

\subsection{The Emergent Physical Picture}

The black hole initially resides in a geometric phase and undergoes Hawking evaporation, during which the heat capacity is negative and the temperature increases as the mass decreases. 
In contrast to the standard Hawking case, the dual structure of the metaparticle—controlled by the duality parameter $\tilde\mu$—prevents the temperature from diverging. 
Instead, the temperature reaches a finite maximum at which the heat capacity diverges, signaling a continuous (second-order) thermodynamic transition. 
At this point, both the entropy and the mass remain finite, and the system undergoes a transition from a geometric to a dual-dominated, non-geometric thermodynamic regime.

This behavior reflects the intrinsically quantum nature of the metaparticle, which must be treated as an entangled object combining geometric and dual degrees of freedom. 
The presence of this entanglement enforces a minimal length scale, or equivalently a minimal horizon area, below which a \emph{purely geometric} thermodynamic description ceases to be valid. 
Evaporation therefore terminates at a finite minimal mass, where the geometric heat capacity becomes positive and vanishes at the endpoint, indicating that further energy exchange through geometric (spacetime) channels is suppressed. 
The final state is thus a cold, stable, non-singular remnant, whose physical characterization requires a non-geometric, modular description \cite{Chouha-PR-cosm:2026-b}.

A potential source of confusion arises from the fact that the dual entropy branch satisfies $S^{-}(\tilde a) > S^{+}(\tilde a)$ for all $\tilde a$. 
This inequality does \emph{not} imply that the dual sector dominates the thermodynamics at all scales. 
Thermodynamic behavior is governed by derivatives of the entropy rather than by its absolute magnitude. 
At large $\tilde a$, both branches share the same leading area-law behavior and identical leading derivatives, rendering the dual microstates thermodynamically inert (c.f. figure \ref{fig:pseudo-entanglement-large-a-loglog}). 
Geometry therefore emerges as an effective description despite the presence of dual degrees of freedom, in direct analogy with the suppression of heavy winding modes in the infrared regime of string theory.

As evaporation proceeds and $\tilde a$ decreases, subleading corrections become increasingly important. 
The geometric entropy branch $S^{+}$ bends downward and vanishes at the minimal area, while the dual branch $S^{-}$ remains smooth and finite. 
At the point of maximal temperature, where the heat capacity changes sign, dual (winding-like) degrees of freedom become thermodynamically active. 
This marks the onset of a non-geometric phase in which a single spacetime polarization can no longer support the thermodynamic description.

\subsection{Constraint-Induced Matter versus Constraint-Induced Geometry}
\label{subsec:constraint_matter_vs_geometry}

A sharp conceptual distinction emerges when comparing mimetic gravity \cite{ChamseddineMukhanov2013} with the metaparticle framework, despite the superficial similarity that both involve constraint-induced reductions of phase space.

\paragraph{Mimetic gravity: emergence of material degrees of freedom.}
In mimetic gravity, the scalar field $\phi$ obeys the constraint
\begin{equation}
g^{\mu\nu}\partial_\mu\phi\,\partial_\nu\phi = 1,
\end{equation}
implemented by a Lagrange multiplier $\lambda$ in the action \cite{ChamseddineMukhanov2013}. 
In the Hamiltonian formulation, the associated constraints are second class, and their elimination leaves behind a genuine physical scalar degree of freedom whose stress--energy tensor
\begin{equation}
T_{\mu\nu}^{\text{mim}} = \lambda\,\partial_\mu\phi\,\partial_\nu\phi
\end{equation}
takes the form of pressureless dust. 
The constraint thus converts a would-be gauge sector into material content.

\paragraph{Metaparticles: emergence of geometry rather than matter.}
By contrast, the metaparticle action (\ref{Eq:Metaparticle}) contains the first-class constraint
\begin{equation}
\varphi_2 := p\cdot\tilde p - \mu \approx 0,
\end{equation}
which generates gauge transformations mixing ordinary and dual coordinates,
\begin{equation}
\delta x^\mu = \varepsilon\,\tilde p^\mu,
\qquad
\delta \tilde x^\mu = \varepsilon\, p^\mu .
\end{equation}
Eliminating $\varphi_2$ removes gauge redundancy rather than producing new propagating degrees of freedom. 
The reduction corresponds to selecting a polarization (or section) of modular spacetime and thereby defines what is meant by physical geometry.

The parameter $\mu$ survives this reduction as a fixed invariant on the constraint surface. 
It does not represent matter or energy, but instead labels inequivalent reduced phase spaces and symplectic structures. 
In this sense, $\mu$ acts as a superselection parameter\footnote{In this context, \emph{superselection} refers to the fact that $\mu$ generalizes the section condition of double field theory or generalized geometry by fixing the embedding of physical momenta within the doubled phase space. Different values of $\mu\in\mathbb{R}$ define inequivalent reduced phase spaces that cannot be coherently superposed.}
analogous to the Immirzi parameter in loop quantum gravity\footnote{More precisely, $\mu$ plays a role analogous to the Immirzi parameter in loop quantum gravity \cite{Rovelli1996Immirzi}, labeling inequivalent quantum realizations of the same classical theory and affecting quantum spectra and entropy formulas \cite{AshtekarBaezKrasnov2000}, while not corresponding to a dynamical degree of freedom.}.

\paragraph{Dust cores versus non-geometric cores.}
This distinction has direct implications for black-hole interiors. 
In mimetic gravity, constraint reduction yields dust-like, material cores \cite{ChamseddineMukhanov2017}. 
In contrast, the metaparticle framework studied here suggests intrinsically non-geometric cores, characterized by regions where spacetime is defined only up to duality transformations c.f.\cite{Chouha-PR-cosm:2026-b} for more details).

\subsection{Minimal Area from the Entropy Reality Condition and its Geometric Interpretation}
\label{subsec:min_area_reality_connection}

We now provide the technical origin of the geometric--to--non-geometric transition described above. 
The entropy reality condition implies the bound
\begin{equation}
\boxed{\tilde a\ge \tilde a_{\min}}
\qquad\Longleftrightarrow\qquad
\boxed{A\ge A_{\min}:=\frac{\pi\,\tilde a_{\min}}{4\tilde\mu},}
\end{equation}
which is not imposed by hand but follows from the requirement that the quantum-corrected entropy remain real.

Because the constraint $\varphi_2$ is first class, its elimination fixes a polarization of modular spacetime rather than generating additional propagating degrees of freedom. 
A thermodynamic interpretation in terms of a geometric horizon is therefore meaningful only insofar as such a geometric polarization exists. 
The entropy reality condition precisely encodes the domain of validity of this geometric description.

As the horizon area approaches $A_{\min}$, the geometric entropy branch $S^{+}$ vanishes, signaling the breakdown of a spacetime thermodynamic interpretation. 
Importantly, this does not imply that thermodynamics itself ceases to apply. 
Rather, the dual entropy branch $S^{-}$ remains finite at $A_{\min}$, indicating that the system transitions to a non-geometric thermodynamic regime dominated by dual (winding-like) degrees of freedom. 
Below $A_{\min}$ no single geometric polarization admits a consistent continuation, but a modular, non-geometric description remains well-defined.

In contrast to mimetic gravity, where regular interiors arise through effective matter sources, the metaparticle framework resolves the endpoint of evaporation through a kinematic obstruction to geometry itself. 
The minimal area thus marks not the disappearance of thermodynamic degrees of freedom, but a phase transition in which geometric entropy is replaced by dual entropy. 
This suggest that the resulting remnant is naturally interpreted as an intrinsically non-geometric core, appropriately described as a T-fold or R-fold configuration patched by duality transformations rather than conventional diffeomorphisms \cite{Hull2005,FreidelLeighMinic2015}.

\subsection{Dual Equation of State and First Hints Towards the Nature of the Remnant Core}
\label{subsec:dual_eos_core}

From a cosmological perspective, as discussed in Section~\ref{Met-MDR}, the modified dispersion relation implies that the geometric (metaparticle-like) branch behaves as radiation with equation of state $w=1/3$, while the dual branch exhibits an effective ultraviolet equation of state
\begin{equation}
w_{\text{dual}} = -\frac{1}{3}.
\end{equation}
At first sight, this value is reminiscent of string winding modes in cosmological settings \cite{brandenberger1989superstrings}. 
In the present framework, however, it should not be interpreted as signaling the emergence of an independent material fluid. 
The dual sector does not correspond to a separate population of propagating degrees of freedom, but rather to the kinematic dominance of the dual component of the same entangled metaparticle once the geometric polarization becomes thermodynamically unstable.
\par An equation of state $w=-1/3$ is characteristic of tension- or curvature-dominated configurations, rather than of particle-like matter. 
Its appearance here should therefore be interpreted as an effective macroscopic signature of the underlying modular structure of spacetime, rather than as evidence for a fundamental fluid description. 
Combined with the entropy reality condition, this is consistent with a picture in which black-hole evaporation terminates at a finite modular scale\footnote{
Here ``modular scale'' refers to the scale at which the description of spacetime in terms of a single geometric polarization breaks down and physical configurations must be identified modulo $O(d,d)$ duality transformations in the doubled (modular) phase space, rather than in terms of a local spacetime metric.
}
, suggesting that the interior may require a non-geometric modular description beyond any single geometric polarization; explicit realizations of this structure will be explored in future work \cite{Chouha-PR-cosm:2026-b}.

\paragraph{Where is the remnant mass stored?}
The remnant mass should be understood as a purely gravitational quantity defined asymptotically, analogous to the ADM mass in general relativity. 
It does not arise from localized energy density but from the global curvature and topological obstruction associated with a nontrivial modular identification of spacetime. 
The entropy reality condition fixes a minimal horizon area and therefore a minimal mass, which represents the gravitational charge required to sustain a non-geometric modular structure\footnote{
By \emph{global curvature} we mean that the spacetime geometry, considered as a whole, is not continuously deformable to flat space, even if no localized curvature singularity is present. The associated gravitational mass is therefore defined asymptotically, through the global behavior of the metric, rather than through a local energy density. 
By \emph{topological obstruction} we mean that the modular identification relating geometric and dual degrees of freedom prevents spacetime from being globally described as a purely geometric manifold. In this sense, the remnant mass reflects the gravitational cost of sustaining a non-geometric (modular) spacetime structure, rather than the presence of material content in the interior.
}
.

\paragraph{Minimal mass as an indirect curvature bound.}
Although no local curvature bound is imposed in the metaparticle framework, the existence of a minimal mass indirectly prevents curvature singularities. 
Singularity avoidance arises not from modified local field equations, but from a kinematic restriction on the class of geometric configurations admitted as physical states. 
Configurations with arbitrarily small horizons are incompatible with the modular consistency conditions of the theory, thereby excluding spacetimes with unbounded curvature. 
The resulting curvature bound is therefore global rather than local, reflecting the modular structure of spacetime rather than dynamical matter content.

For clarity, Table~\ref{tab:mimetic_vs_metaparticle_remnants} summarizes the similarities and key distinctions between the non-singular mimetic black hole and our metaparticle-corrected black hole remnant.
\begin{table}[H]
\centering
\renewcommand{\arraystretch}{1.25}
\begin{tabular}{|p{4.2cm}|p{5.4cm}|p{5.4cm}|}
\hline
\textbf{Aspect} 
& \textbf{Mimetic gravity non-singular remnant} 
& \textbf{Metaparticle non-singular remnant} \\
\hline
Underlying mechanism 
& Local dynamical constraint on the metric enforced by a scalar field 
& First-class constraint selecting a polarization of modular (doubled) spacetime \\
\hline
Type of constraint 
& Second-class constraints; elimination leaves physical degrees of freedom 
& First-class constraint; elimination removes gauge redundancy \\
\hline
What the constraint produces 
& Effective material degree of freedom (mimetic dust) 
& Definition of geometry itself (choice of section/polarization) \\
\hline
Nature of the remnant core 
& Matter-filled, dust-like, local and extensive 
& Non-material, non-geometric, modular (duality-defined) core (T-fold/ R-fold type)\\
\hline
Origin of non-singularity 
& Explicit local bound on curvature invariants 
& Global obstruction to extending a geometric description below a minimal modular scale \\
\hline
Curvature resolution 
& Curvature bounded dynamically everywhere 
& Curvature bounded indirectly via minimal horizon area \\
\hline
Minimal scale 
& Set by curvature bound in field equations 
& Set by entropy reality condition ($A \ge A_{\min}$) \\
\hline
Remnant mass 
& Associated with effective dust energy density 
& ADM gravitational charge of a topological/modular defect \\
\hline
Where the mass is ``stored'' 
& In matter-like stress--energy of mimetic dust 
& In global geometry and boundary conditions, not local energy density \\
\hline
Thermodynamic behavior near endpoint 
& Maximal temperature, finite entropy, stable remnant 
& Maximal temperature, finite entropy, stable remnant \\
\hline
Heat capacity at endpoint 
& Changes sign; stabilizes remnant 
& Changes sign; goes to $0^{+}$ at the minimal mass \\
\hline
Effective UV equation of state 
& Dust-like ($w=0$) in the core 
& Tension/curvature-like ($w=-1/3$) from dual sector dominance \\
\hline
Interpretation of UV degrees of freedom 
& Material (effective cold dark matter) 
& Geometric/topological (dual, winding-like, non-local) \\
\hline
Geometric description of the core 
& Regular spacetime with matter content 
& Breakdown of single geometric polarization; duality-patched spacetime \\
\hline
Conceptual role of the remnant 
& Matter-supported regular black hole core 
& Frozen modular/topological defect of spacetime (with finite non-extensive entropy) \\
\hline
\end{tabular}
\caption{Comparison between non-singular black-hole remnants in mimetic gravity and in the metaparticle framework. Although both scenarios yield finite-temperature endpoints and stable remnants, the microscopic origin and physical interpretation of the remnant differ fundamentally: mimetic gravity resolves singularities through emergent matter and local curvature bounds, while the metaparticle framework resolves them through modular geometry, entropy reality conditions, and global topological obstructions.}
\label{tab:mimetic_vs_metaparticle_remnants}
\end{table}
Beyond the comparison with mimetic gravity, it is instructive to contrast the metaparticle remnant with several other proposals for non-singular black holes.

Non-singular black-hole remnants arise in a variety of approaches to quantum and modified gravity, but the physical nature of the resulting cores differs markedly across frameworks. 
In several theories, singularity resolution is achieved through the emergence of effective matter sources in the interior. 
This occurs most explicitly in mimetic gravity (as discussed above). 
Related induced-matter interiors also appear in limiting-curvature models \cite{MukhanovBrandenberger1992,ChamseddineMukhanov2017LC}, Palatini $f(R)$ gravity \cite{OlmoRubieraGarcia2015}, loop-quantum-gravity--inspired black holes \cite{AshtekarBojowald2006,AshtekarOlmedoSingh2018}, and asymptotically safe gravity \cite{BonannoReuter2000}, where high-curvature corrections generate effective stress--energy components that regularize the core.

A conceptually distinct class of resolutions replaces the singularity by a transition to a white-hole geometry or a bounce into an expanding interior region, as in black-to-white-hole scenarios motivated by loop quantum gravity or effective polymer quantization \cite{RovelliVidotto2014,HaggardRovelli2015,ChristodoulouRovelli2015}. 
In these models, the endpoint is not a static remnant but a dynamical spacetime continuation, often accompanied by a temporary trapping horizon.

A further class of non-singular black-hole models arises from non-commutative geometry approaches, in which spacetime coordinates satisfy nontrivial commutation relations that effectively smear point-like sources over a minimal length scale. 
In these models, classical singularities are resolved because the energy--momentum tensor is replaced by a non-local, Gaussian-smeared distribution, leading to regular black-hole solutions with de~Sitter--like or finite-density cores \cite{Nicolini:2005vd,Spallucci:2008ez}. 
While non-commutative black holes share with the present framework the existence of a minimal length and a remnant mass, the underlying mechanism is fundamentally different: non-commutative models resolve singularities through non-local matter smearing in a fixed geometric background, whereas the metaparticle framework replaces the notion of a smooth interior altogether by a modular, non-geometric core determined by entropy reality and duality constraints, rather than by an effective stress--energy distribution.

Finally, as mentioned in section \ref{Reg-BH}, singularity avoidance can occur without horizons or interior matter, leading to horizonless compact objects such as gravastars \cite{MazurMottola2004}, fuzzballs \cite{Mathur2005,Mathur2009}, and other exotic compact objects \cite{CardosoPani2019}. In these cases, the classical notion of a black-hole interior is replaced altogether by nonperturbative microstructure or boundary-layer physics.
The metaparticle framework considered here belongs to none of these categories. 
Its non-singular remnant does not arise from emergent matter, white-hole continuation, or horizonless microstructure, but instead follows from a kinematic restriction imposed by modular spacetime together with the entropy reality condition. 
As established above, the resulting core is a finite, non-geometric modular core, carrying a nonzero ADM mass interpreted as a global gravitational charge rather than as localized energy density. 
In this sense, the remnant is naturally interpreted as a topologically protected defect of modular spacetime, whose detailed structure remains an interesting direction for future investigation \cite{Chouha-PR-cosm:2026-b}.

\section{Conclusion and Discussion}

In this work we have studied black hole thermodynamics and evaporation within the framework of metastring theory, using metaparticles as the relevant quantum probes of spacetime at high energies. Metaparticles encode intrinsic UV/IR mixing through a duality constraint relating ordinary and dual momenta, and therefore provide a concrete setting in which quantum-gravitational effects enter black hole physics beyond effective field theory.

Starting from a generalized Bekenstein argument adapted to the metaparticle dispersion relation, we identified two entropy contributions associated with geometric (momentum-like) and dual (winding-like) sectors of the underlying modular phase space. When these sectors are treated as effectively independent or additive, the resulting thermodynamics is incomplete and leads to unphysical behavior at small horizon area, including negative entropy and inconsistencies between temperature and heat capacity. This signals a breakdown of the independent-sector approximation rather than a failure of the underlying framework.

By treating the metaparticle as a genuinely entangled quantum object, we proposed a pseudo-entangled total entropy that incorporates correlations between the two sectors. A key result is that the requirement of entropy reality dynamically enforces a minimal horizon area and, equivalently, a minimal black hole mass. This minimal scale is not imposed by hand and does not arise from modified field equations or matter sources, but follows directly from the structure of the entropy itself. As a consequence, black hole evaporation proceeds through a finite maximal temperature, undergoes a continuous (second-order) thermodynamic phase transition, and Hawking radiation shuts down leaving behind a cold, stable remnant in which geometric thermodynamics channels are inactive.
\par The resulting remnant differs qualitatively from those appearing in many other approaches to singularity resolution. 
In particular, it is not supported by emergent matter degrees of freedom, as in mimetic gravity or induced-matter models, nor by local curvature bounds imposed at the level of the equations of motion. 
Instead, the remnant is naturally interpreted as a finite modular core of spacetime, where a single geometric polarization ceases to be globally valid and the interior is more appropriately described as a non-geometric, duality-patched region. 
From this perspective, the remnant can be viewed as a topologically protected defect of modular spacetime: its existence is enforced by global consistency conditions—most notably the duality constraint and the requirement of entropy reality—rather than by local dynamical stabilization. 
The remnant mass is therefore a global gravitational charge, analogous to an ADM invariant, rather than the integral of a local energy density.


From a broader perspective, our results suggest that entropy reality conditions and modular spacetime structure can play a central role in determining the endpoint of black hole evaporation. The metaparticle framework provides a concrete realization of this idea, in which the interplay between geometry, duality, and quantum correlations naturally leads to a stable, non-singular endpoint without introducing exotic matter or violating semiclassical consistency conditions such as adiabaticity. In this sense, black hole evaporation does not terminate in a material core, but in a topologically protected obstruction to a purely geometric description of spacetime.

\par Beyond its implications for black hole evaporation, the present analysis clarifies the physical role of several key ingredients of metastring theory and its metaparticle excitations. In particular, we see that the first-class diffeomorphism constraint of the metaparticle theory is not merely a formal consistency condition, but enforces a nontrivial relation between particle and dual sectors that prevents the physical Hilbert space from factorizing. As a result, the metaparticle and its dual are necessarily entangled, giving rise to an intrinsic form of non-locality that links geometric and dual (winding-like) degrees of freedom associated with different spacetime polarizations. This constraint-induced non-locality obstructs the existence of a globally valid local geometric description of spacetime. In the black hole context, this obstruction becomes manifest through the requirement of entropy reality, which enforces a minimal horizon area, halts evaporation at a finite mass, and leads to a stable remnant naturally interpreted as a topologically protected defect of modular spacetime. 

\par Put differently, at the endpoint, entropy no longer counts spacetime configurations but instead reflects a small, finite degeneracy of the remaining non-geometric quantum state enforced by the constraints\footnote{
The finite value of the dual entropy at the endpoint should not be interpreted as the logarithm of a large number of independent spacetime microstates. 
In the non-geometric regime, entropy is non-extensive and reflects a residual, constraint-induced degeneracy of the physical Hilbert space rather than a count of localized degrees of freedom. 
As a result, its numerical value may be of order unity or smaller without signaling any inconsistency.
}.

Several directions for future work are open. A microscopic derivation of the correlation entropy from an explicit density-matrix construction would further sharpen the interpretation of the pseudo-entanglement term. It would also be interesting to investigate potential observational signatures of modular remnants, for instance, through gravitational-wave echo phenomena associated with nonstandard near-horizon or interior structure \cite{Cardoso:2016rao,Cardoso:2019rvt}. In a cosmological context, the non-thermal and topologically protected nature of the remnant suggests possible connections to memory-burden effects \cite{Dvali:2020wqi} and to imprints on early-universe observables, such as the cosmic microwave background, if such objects form primordially or influence late-time entropy production \cite{Carr:2020gox}. More broadly, the present analysis suggests that modular spacetime and duality constraints may offer a unifying language for understanding black hole interiors, singularity resolution, and the quantum structure of spacetime itself.

\section*{Appendices}
\appendix

\section{The Original Bekenstein Argument: with unmodified dispersion relation and uncertainty principle}\label{SubSec:Original-Bek}
We want to start by recalling (following \cite{Bousso:2002ju}) the derivation of the general relativistic result that the minimum increase in the area of a black hole when it absorbs  classical matter of energy $E$ and size $R$ is given by
\begin{equation}\label{Area-Bek-cl}
    \Delta A\simeq 8\pi L_{P}^{ 2}\,  \delta E\, R
\end{equation}
in units where $(\hbar=c=1)$ which we will recover later.
$\delta E$ is the total energy of a small unit of matter located at a proper distance $R$ of the horizon.

Since we want to compute the minimal possible increase in the black hole's horizon area resulting from capturing that matter system, we will assume a {\it{Geroch process}}\footnote{The process was presented in 1972 in a colloquium at Princeton university, and suggests that all of the energy of a certain system can be extracted by lowering it into a black hole.  } and not let the system fall from infinity (since this will increase the area by exactly $\delta E$).\\
\par The idea is to extract work from the system by lowering it slowly until it is just outside the event horizon of the black hole and then we let it go. Since the system has a center of mass (CM) (and is not an idealized point particle), the mass added to the system is redshifted according to the position of the CM at the drop-off point (the point where the circumscribing sphere is taken to almost touch the horizon). We can orient the CM so that it is at a proper distance $R$ from the horizon. Classically, the reason why we cannot let $R=0$ (i.e., position the CM at the horizon) is that it would imply that part of the system is past the event horizon and tidal forces would rip apart the entire system. Quantum mechanically, Heisenberg's uncertainty relation will prevent us from taking $R=0$ (as we will see below). Thus, one must compute the redshift factor at radial proper distance $R$ from the horizon.\\
Our starting point is the Schwartzschild metric 
\begin{equation}\label{Schwartzchild-metric}
ds^2=-f(r)dt^2+f(r)^{-1}dr^2+r^2d\Omega_2     
\end{equation}
where $f(r)$ defines the redshift factor $\alpha_g(r)$
\begin{equation}\label{redshift-factor}
    f(r)=1-\frac{r_H}{r}\equiv \alpha_g^{2}(r)
\end{equation}
and $r_H=2GM/c^2$ is the Schartzschild radius of the black hole and $M$ its mass measured at infinity.\\
Let $h$ denote the radial coordinate distance from the horizon
\begin{equation}
    h=r-r_H
\end{equation}
Near the horizon, the square of the  redshift factor $f(h)$ (to leading order in $h/r_H$) is given by
\begin{equation}
    f(h)=1-\left(1+\frac{h}{r_H}\right)^{-1}\approx \frac{h}{r_H}.
\end{equation}
The proper distance $l$ is related to the coordinate distance $h$ by
\begin{equation}\label{proper-dist-sch}
    l(h)=\int_{0}^{h}\frac{dh}{\sqrt{f(h)}}=2\sqrt{hr_H}.
\end{equation}
Hence, it terms of the proper distance $l$, the redshift factor reads
\begin{equation}\label{red-shift-l}
    \alpha_g(l)\equiv\sqrt{f(l)}\approx \frac{l}{2\, r_H}
\end{equation}
Therefore, the mass added to the black hole is
\begin{equation}\label{change-in-mass}
    \delta M c^2= \Gamma\delta E\  \alpha_g(l) \left.\right\vert_{l=R}\approx \Gamma  
 \ \frac{\delta E\, R}{2\, r_H}.
\end{equation}
To translate the bound into the change in the area of the horizon we just use $A=16\pi G^2 M^2/c^4$ or $\delta A=16\pi Gr_h \delta M/c^2$ together with equation (\ref{change-in-mass}) and thus the increase in the area of the Black hole's horizon is given by 
\begin{equation}\label{Bekenstein-Area-bound}
    \delta A\simeq \frac{8\pi G}{c^4} \Gamma \delta E\, R
\end{equation}
which is equation (\ref{Area-Bek-cl}) above, where we have introduced back all constants to emphasize that the classical expression does not involve any $\hbar$ nor Planck constants. Note, that we introduced a calibration factor $\Gamma$ such that, classically, $\Gamma=1$, while $\Gamma\neq 1$ otherwise.
\par To describe the absorption of  {\bf{quantum}} matter of rest-mass $M$, one must describe the CM position of the spherical matter in terms of Heisenberg's uncertainty in its position, $R\sim\delta x$.
To measure its position, one must shine say on the CM of the system a photon of energy $E_\gamma$ and momentum $p_\gamma$. According to Heisenberg's UP, the photon must have a momentum $\delta p_\gamma\geqslant \hbar/2\delta x$. Then one uses the special relativistic dispersion relation to relate the energy of the photon to its momentum $\delta E_\gamma=c\delta p_\gamma \geqslant \hbar c/2\delta x$.
\paragraph\
To ensure that the relevant energy uncertainties are not large enough to allow the production of additional copies of the quantum matter whose position we are trying to measure, the measurement procedure requires that $\delta E_\gamma\leqslant Mc^2$. Therefore, the relation $\delta E_\gamma\geqslant \hbar c/2\delta x$ is converted to $Mc^2\geqslant \hbar c/2\delta x$. Although the procedure outlined above applies to a system at rest, one can generalize the result (for example by using a boost) to the case of a system whose total energy is $E$ \cite{Lifshitz} and thus we get
\begin{equation}\label{eq:EnergyUP}
   \delta E\geqslant \hbar c/2\delta x. 
\end{equation}
 Putting all the above together into equation (\ref{Area-Bek-cl}), we get
\begin{equation}\label{Quantum-Area-Ent-Bek}
\Delta A \simeq \frac{8\pi G\, \Gamma}{c^4} \delta E\delta x.
\end{equation}

For this scenario, we will fix the calibration factor $\Gamma$ by 
 connecting the classical result (\ref{Area-Bek-cl}) to its quantum counterpart which leads to
\begin{equation}\label{Area-Bek-quantum-Fin}
    \Delta A \geqslant \frac{8\pi G \Gamma}{c^4} \frac{\hbar c}{2}=\frac{4\pi\Gamma \hbar G}{c^3}=4\pi\Gamma L_P^2 \end{equation}
\par
By Landauer's principle \cite{landauer1961irreversibility}, the loss (erasure) of one bit of information requires a minimum energy cost of $k_B T \ln{2}$ which would correspond to a minimal increase in entropy $\Delta S_{min}=k_B\ln{2}$.
Thus, the minimal change in entropy accompanying the minimal change in the area of the event horizon of the black hole due to the absorption of the quantum matter is the reciprocal of
\begin{equation}\label{Area-Entropy-Unmod}
   \frac{(\Delta A)_{\text{min}}}{(\Delta S)_{\text{min}}}\simeq \frac{dA}{dS}\simeq \frac{4\pi\Gamma L_P^2} {k_b\ln2}
\end{equation}
and thus 
\begin{equation}\label{Entropy-Area-Unmod}
    \frac{dS}{dA}\simeq \frac{k_B\ln2}{4\pi\Gamma L_P^2}
\end{equation}
which upon integration and setting $\Gamma\equiv \ln2/\pi$ leads to the famous linear dependence of the Entropy of the black hole on its horizon's area
\begin{equation}\label{BH-Ent-Are-orig}
    S=\frac{k_B A}{4 L_P^2}.
\end{equation}
Our starting point will be equation (\ref{Bekenstein-Area-bound}) in the form
\begin{equation}\label{Main-Area-Calib}
\Delta A\simeq \frac{8\pi\Gamma L_P^2 E\delta x}{\hbar c}  
\end{equation} where $\Gamma$ is a calibration factor that we will fix later. In natural units we will write
\begin{equation}\label{Eq:Area-Plnakc-units}
    \Delta\tA\simeq 8\pi\,\Gamma \tE\,\delta \tx 
\end{equation}
where all the tilde quantities are with respect to their Planckian values (for ex. $\tA=\frac{A}{L_p^2}$ etc.) 
\section{Modular Spacetime}\label{ModST}

In this appendix we expand on the notions introduced in the main text. One of the
central challenges in quantum gravity (QG) is to construct a theory that reconciles
the existence of a fundamental length scale, such as the Planck length, with Lorentz
invariance. Approaching the problem from this perspective, recent work
\cite{Freidel:2015uug,Freidel:2016pls,Freidel:2017xsi} introduced a new notion of
quantum spacetime called \emph{modular spacetime}, building on the ideas of
\cite{Aharonov:1969qx} that quantum observables are modular, such as
$([q]_R,[p]_{2\pi\hbar/R})$, where
$[q]_R:=q\!\mod R$ and $[p]_{2\pi\hbar/R}:=p\!\mod\!\left(\frac{2\pi\hbar}{R}\right)$.
We identify $[q]:=[x]$ and $[p]:=[\tilde{x}]$, where $R$ sets the size of the
fundamental modular cell. The observables $[\tilde{x}]$ have no classical analogue,
since they do not survive the limit $\hbar\to0$.

Quantum (modular) spacetime is therefore defined as the maximal set of commuting
modular observables. The classical continuum spacetime is recovered in the limit
$R\to\infty$ \cite{Freidel:2015pka}. This construction may be viewed as a phase-space
formulation related to doubled geometric frameworks, defined on a self-dual lattice
$l\oplus\tilde l$, naturally giving rise to generalized objects
$\mathbb{X}^A:=([x]^a,[\tilde{x}]_a)^T$. The resulting doubly orthogonal structure is
captured by a neutral $O(D,D)$ metric $\eta$.

The simplest geometric structure arising from the modular treatment of spacetime is
the doubled line element
\begin{equation}\label{Eq:metric}
ds^2=g_{\mu\nu}dx^{\mu}dx^{\nu}+g^{\mu\nu}d\tilde{x}_{\mu}d\tilde{x}_{\nu},
\end{equation}
together with the duality pairing
\begin{equation}\label{Eq:duality-metric}
\Theta = dx^{\mu}d\tilde{x}_{\mu}.
\end{equation}

The metaparticles can be thought of the excitations of such a quantum system \cite{Freidel:2018apz}.

\section{Metastrings and New Roles of Old Fields}\label{MString}
It turns out that this mathematical structure (and more precisely this $\eta\in O(D,D)$ metric) captures the idea of Born reciprocity \cite{Born1938ASF} which unifies the structure of spacetime with that of energy-momentum space. In that sense, what we now call spacetime is probe dependent, a principle the authors of \cite{Amelino-Camelia:2011lvm} dubbed {\it{relative locality}} which seems to be the natural organizational principle underlying string theory. The $\eta$-metric is a polarization metric which tells us if we are spacetime-like or momentum-energy-like (or anywhere in between) in this new $2D$ phase space. 
\par The discretized nature of phase-space is captured by a natural symplectic structure (the two-form) $\omega\in Sp(2D)$ associated with the commutator. Furthermore, in order to define a vacuum (since we are in a discretized-lattice setting given that we cannot apply the translational generators on the vacuum), we need a mathematical object which is invariant with respect to the previous two structures $\eta$ and $\omega$. This object is a conformal double metric that is called the {\it{quantum metric}} $H$ which is an $O(2,2(D-1))$\footnote{The first 2 in the $O(2,2(D-1))$ does not refer to having two time directions but refers to a time-like direction and to an energy-like direction (which completes the realization of the ideas of Born-reciprocity \cite{Born1938ASF} and relative locality \cite{Amelino-Camelia:2011lvm} i.e.\ that different observers experience different spacetimes that is different slices of modular-spacetime).} object sometimes named the generalized metric in the context of DFT or Generalized Geometry.\\
Thus to recap, the three structures present are a symplectic two-form $\omega$,
a neutral $O(D,D)$ metric $\eta$, and a double quantum metric $H$.
The compatibility of these structures selects the Lorentz subgroup $O(D)$.
\par Together, the compatible triple $(\eta,\omega,H)$ defines the Born geometry of
modular spacetime \cite{FreidelLeighMinic2015}, encoding respectively the
duality pairing, the symplectic phase-space structure, and the generalized
quantum metric.

\par These three structures arise naturally in string theory when the Tseytlin action
\cite{Tseytlin:1990nb} is written as:
\begin{equation}\label{Eq:Modified-Tseytlin}
    S_{Metastrings}=\frac{1}{4\pi}\int_{0}^{2\pi}d\sigma\int_{\tau_i}^{\tau_f}d\tau\left(\partial_{\tau}\mathbb{X}^A\left(\eta_{AB}+\omega_{AB}\right)\partial_{\sigma}\mathbb{X}^B-\partial_{\sigma}\mathbb{X}^A H_{AB} \partial_\sigma\mathbb{X}^B\right)
\end{equation}
where the notion of coordinates is generalized to a $2D$-vector $\mathbb{X}^M=\begin{pmatrix}
x^{\mu}\\ \tilde{x}_{\mu}\\

\end{pmatrix}$.\\
The {\bf{metastring}} introduced in \cite{Freidel:2015pka} and \cite{Freidel:2017xsi} is defined by the above action (\ref{Eq:Modified-Tseytlin}).
It is important to note that the new term $\omega_{AB}$ is usually disregarded in the formulation of the string action since it is only viewed as a topological term. It turns out that this term plays a very important role in the geometrical structure of the metastring theory and sheds important new light on the role of the Kalb-Ramond field.
In the absence of the Kalb-Ramond field $B$, the triple-$(\eta^{(0)}, \omega^{(0)}, H^{(0)})$ are given by:
\begin{equation}\label{Eq:Eta-Omega-H-zeroB}
\begin{aligned}
    &\eta_{AB} =\begin{pmatrix}
0 & \delta_{i}^{\ j}\\
\delta^{i}_{\ j} & 0
\end{pmatrix},\  \omega_{AB}=\begin{pmatrix}
0 & -\delta_{i}^{\ j}\\
\delta^{i}_{\ j} & 0
\end{pmatrix},&\\
&\text{and\ } H_{AB}=\begin{pmatrix}
g_{ij} & 0 \\
0 & g^{ij}
\end{pmatrix}.&
\end{aligned}
\end{equation}
Note that in DFT and Generalized Geometry $H_{AB}$ mixes both the Kalb-Ramond field $B$ and the metric $G$ 
\begin{equation}\label{Eq:Gen-metric}
    H_{DFT}=\begin{pmatrix}
G-BG^{-1}B & BG^{-1}\\
-G^{-1}B & G^{-1}
\end{pmatrix}
\end{equation}
In the metastring formulation in the presence of the Kalb-Ramond field $B$, only the 
two-form $\omega^{(B)}$ is modified and given by:

\begin{equation}\label{Eq:Eta-Omega-H-B}
\begin{aligned}
    \omega_{AB}^{(B)}=\begin{pmatrix}
-2B_{ij} & -\delta_{i}^{\ j}\\
\delta^{i}_{\ j} & 0
\end{pmatrix},
\end{aligned}
\end{equation}

Thus we see that the $B$-field is not contained in the generalized metric
$H_{AB}$ but rather in the symplectic two-form $\omega$, providing a distinct
geometrical interpretation of the role of the Kalb--Ramond field.
In light of this, we can consider the {\bf{$B$-transformation}}:
\begin{equation}\label{Eq:B-transformation}
\mathbb{X}:=(x^i,\tilde{x}_i)^T\xrightarrow{B}(x^i,\tilde{x}_i+B_{ij}x^j)^T,
\end{equation}
which gives us the commutation relation on the generalized phase space 
$[\mathbb{X}^A,\mathbb{X}^B]=2\pi i\lambda^2(\omega^{(B)})^{AB}$ which leads to the non-commutativity of the dual-space coordinates
\begin{equation}\label{Eq:Non-comm-dual-coord}
    [\tilde{x}_i,\tilde{x}_j]=-4\pi i \lambda^2 B_{ij},
\end{equation}
where $\lambda$ is the length scale of the theory which is related to the energy scale $\epsilon$ via: $\alpha^{\prime}=\lambda/\epsilon$. 
More generally, we can parametrize an arbitrary $O(D,D)$ transformation by nilpotent transformations 
$h=e^{\hat{B}}\hat{A}e^{\hat\beta}$ where $\hat{A}\in GL(D)$ and
\[
e^{\hat{B}}=
\begin{pmatrix}
1 & 0\\
B & 1
\end{pmatrix},
\qquad
e^{\hat{\beta}}=
\begin{pmatrix}
1 & \beta\\
0 & 1
\end{pmatrix}.
\]

Similarly,we can consider the {\bf{$\beta$-transformation}}: 
\begin{equation}\label{Eq:beta-transformation}
\mathbb{X}:=(x^i,\tilde{x}_i)^T\xrightarrow{\beta}(x^i+\beta^{ij}\tilde{x}_j,\tilde{x}_i)^T,
\end{equation}
which leads to the non-commutativity of the spacetime coordinates 
\begin{equation}\label{Eq:Non-com-space-coord}
    [x^i,x^j]=4\pi i \lambda^2 \beta^{ij}. 
\end{equation}
Thus $\beta$-field backgrounds are non-geometric and non-local. 
\par To summarize, the metastring provides a manifestly T-dual, phase-space
formulation of string theory characterized by the additional symplectic
two-form $\omega$, which constitutes the key structural ingredient beyond the
usual Polyakov formulation. The presence of the Kalb--Ramond field within the
symplectic structure, together with its T-dual $\beta$-field, leads respectively
to the non-commutativity of dual spacetime coordinates and of spacetime
coordinates themselves. In conventional Double Field Theory (DFT), by contrast,
the doubled coordinates $(x,\tilde{x})$ are treated as commuting classical
variables, yielding a doubled geometric description but not an intrinsically
non-commutative doubled spacetime. The metastring framework therefore naturally
realizes a doubled \emph{phase-space} geometry in which non-commutativity is built
into the kinematics of spacetime itself.

\section{Adiabaticity and Back-reaction}
\label{App:Adiabaticity}

A standard consistency requirement for semiclassical black-hole evaporation is that the background geometry evolve slowly compared to the horizon light-crossing time \cite{Page1976,Hiscock1981,Parentani1994,FabbriNavarro}. 
In Planck units ($G=c=\hbar=k_B=1$), the characteristic dynamical time scales as
\[
t_{\rm dyn}\sim t_{\rm cross}\sim \mathcal O(M).
\]
Quasi-static evolution therefore requires the mass-change timescale
\[
t_M \sim \frac{M}{|\dot M|}
\]
to satisfy
\begin{equation}
\frac{t_{\rm dyn}}{t_M}
\sim
\frac{M}{M/|\dot M|}
=
|\dot M|
\ll 1,
\label{Eq:AdiabaticityCondition}
\end{equation}
so that the fractional mass change per crossing time remains small.

We model the luminosity using a Stefan--Boltzmann law,
\begin{equation}
P=\gamma\,g\,\sigma\,A\,T^4,
\end{equation}
where $g$ is the effective number of radiated relativistic degrees of freedom, 
$\sigma=\pi^2/60$ is the Stefan--Boltzmann constant, and $\gamma\le1$ is an effective greybody factor. 
Frequency-dependent greybody corrections would only reduce the luminosity and thus strengthen the bounds derived below.

Using $A=16\pi M^2$ together with the rescalings
\[
M=\frac{\tilde m}{\sqrt{\tilde\mu}},
\qquad
T=\sqrt{\tilde\mu}\,\tilde\tau(\tilde m),
\]
one obtains
\begin{equation}
|\dot M|
=
16\pi\,\gamma\,g\,\sigma\,
\tilde\mu\,
\tilde m^{2}\tilde\tau(\tilde m)^{4}.
\label{Eq:dotM_rescaled}
\end{equation}
The quasi-static condition \eqref{Eq:AdiabaticityCondition} therefore becomes
\begin{equation}
16\pi\,\gamma\,g\,\sigma\,
\tilde\mu\,
\max_{\tilde m\ge\tilde m_{\min}}
\!\left(\tilde m^{2}\tilde\tau(\tilde m)^{4}\right)
\ll 1.
\label{Eq:mu_bound_general}
\end{equation}

\subsection*{Conservative bound on $\tilde\mu$}

The function $\tilde m^{2}\tilde\tau(\tilde m)^4$ reaches its maximum near the temperature peak at $\tilde m=\tilde m_*$. 
Using the numerical values
\[
\tilde m_*\simeq0.311,
\qquad
\tilde\tau_{\max}\simeq0.098,
\]
one finds
\[
\tilde m_*^2\,\tilde\tau_{\max}^4
\simeq
8.9\times10^{-6},
\]
and hence
\begin{equation}
|\dot M|_{\max}
\simeq
(7.4\times10^{-5})\,g\,\tilde\mu.
\end{equation}
Adiabaticity is therefore guaranteed provided
\begin{equation}
\tilde\mu
\ll
\frac{1}{(7.4\times10^{-5})\,g}.
\label{Eq:mu_bound_numeric}
\end{equation}
For $g=106.75$ (all Standard Model species relativistic) this gives
$\tilde\mu\ll1.3\times10^{2}$, while even an ultra-conservative choice $g\sim10^3$ yields
$\tilde\mu\ll1.4\times10^{1}$.
The value used in this work lies safely within this quasi-static regime.

\subsection*{Endpoint behaviour and validity of the quasi-static approximation}

The strongest constraint arises near the luminosity maximum, i.e. around $\tilde m\simeq\tilde m_*$. 
By contrast, the late stages of evaporation are dynamically milder. 
As the remnant configuration is approached,
\[
\tilde\tau(\tilde m)\rightarrow0
\qquad
(\tilde m\rightarrow\tilde m_{\rm rem}),
\]
so that
\begin{equation}
|\dot M|
\propto
\tilde m^2\tilde\tau(\tilde m)^4
\longrightarrow 0.
\end{equation}
The luminosity therefore vanishes at the endpoint, and the evaporation rate decreases continuously until the remnant is reached. 
The quasi-static approximation thus becomes increasingly accurate in the final stage, in contrast with standard Hawking evaporation where the temperature grows and back-reaction eventually becomes large.

Overall, back-reaction effects remain small throughout the evolution, supporting a self-consistent quasi-static thermodynamic description of the evaporation process.

\section*{Acknowledgments}

The author would like to thank Robert Brandenberger for valuable discussions and for his insightful comments on the manuscript. Special thanks goes also to Heliudson Bernardo for the long fruitful discussions and his encouragement.
\newpage 
\phantomsection
\addcontentsline{toc}{section}{References}

\let\oldbibliography\thebibliography
\renewcommand{\thebibliography}[1]{
  \oldbibliography{#1}
  \setlength{\itemsep}{0pt} 
  \footnotesize 
}
\bibliographystyle{bibstyle} 
\bibliography{References}

@article{Amelino-Camelia:1999hpv,
    author = "Amelino-Camelia, Giovanni",
    editor = "Kowalski-Glikman, J.",
    title = "{Are we at the dawn of quantum gravity phenomenology?}",
    eprint = "gr-qc/9910089",
    archivePrefix = "arXiv",
    reportNumber = "CERN-TH-99-223",
    journal = "Lect. Notes Phys.",
    volume = "541",
    pages = "1--49",
    year = "2000"
}

@article{Adler:2001vs,
  author        = "Adler, Ronald J. and Chen, Pisin and Santiago, David I.",
  title         = "{The Generalized uncertainty principle and black hole remnants}",
  journal       = "Gen. Rel. Grav.",
  volume        = "33",
  pages         = "2101--2108",
  year          = "2001",
  eprint        = "gr-qc/0106080",
  archivePrefix = "arXiv",
  reportNumber  = "SLAC-PUB-8853",
  doi           = "10.1023/A:1015281430411"
}

@article{Adler:2010wf,
  author        = "Adler, Ronald J.",
  title         = "{Six easy roads to the Planck scale}",
  journal       = "Am. J. Phys.",
  volume        = "78",
  pages         = "925--932",
  year          = "2010",
  eprint        = "1001.1205",
  archivePrefix = "arXiv",
  primaryClass  = "gr-qc",
  doi           = "10.1119/1.3439650"
}

@article{Aharonov:1969qx,
  author        = "Aharonov, Y. and Petersen, A. and Pendleton, H.",
  title         = "{Modular variables in quantum theory}",
  journal       = "Int. J. Theor. Phys.",
  volume        = "2",
  pages         = "213--230",
  year          = "1969",
  doi           = "10.1007/BF00670008"
}

@article{Amelino-Camelia:2011lvm,
  author        = "Amelino-Camelia, Giovanni and Freidel, Laurent and Kowalski-Glikman, Jerzy and Smolin, Lee",
  title         = "{The principle of relative locality}",
  journal       = "Phys. Rev. D",
  volume        = "84",
  pages         = "084010",
  year          = "2011",
  eprint        = "1101.0931",
  archivePrefix = "arXiv",
  primaryClass  = "hep-th",
  doi           = "10.1103/PhysRevD.84.084010"
}

@article{AshtekarBojowald2006,
  author        = {Ashtekar, Abhay and Bojowald, Martin},
  title         = {Quantum Geometry and the Schwarzschild Singularity},
  journal       = {Classical and Quantum Gravity},
  volume        = {23},
  pages         = {391--411},
  year          = {2006},
  eprint        = {gr-qc/0509075},
  archivePrefix = {arXiv},
  doi           = {10.1088/0264-9381/23/2/008}
}

@article{AshtekarOlmedoSingh2018,
  author        = {Ashtekar, Abhay and Olmedo, Javier and Singh, Parampreet},
  title         = {Quantum Extension of the Kruskal Spacetime},
  journal       = {Physical Review D},
  volume        = {98},
  number        = {12},
  pages         = {126003},
  year          = {2018},
  eprint        = {1806.02406},
  archivePrefix = {arXiv},
  primaryClass  = {gr-qc},
  doi           = {10.1103/PhysRevD.98.126003}
}

@article{Bardeen1968,
  author        = {J. M. Bardeen},
  title         = {Non-singular black hole},
  journal       = {Nature},
  volume        = {182},
  pages         = {147-148},
  year          = {1968},
  doi           = {10.1038/182147a0}
}

@article{Bekenstein:1973ur,
  author        = {Bekenstein, Jacob D.},
  title         = {Black holes and entropy},
  journal       = {Phys. Rev. D},
  volume        = {7},
  pages         = {2333--2346},
  year          = {1973},
  doi           = {10.1103/PhysRevD.7.2333}
}

@article{Bekenstein:1974ax,
  author        = {Bekenstein, Jacob D.},
  title         = {Generalized second law of thermodynamics in black-hole physics},
  journal       = {Phys. Rev. D},
  volume        = {9},
  pages         = {3292--3300},
  year          = {1974}
}

@article{BonannoReuter2000,
  author        = {Bonanno, Alfio and Reuter, Martin},
  title         = {Renormalization Group Improved Black Hole Spacetimes},
  journal       = {Physical Review D},
  volume        = {62},
  pages         = {043008},
  year          = {2000},
  eprint        = {hep-th/0002196},
  archivePrefix = {arXiv},
  doi           = {10.1103/PhysRevD.62.043008}
}

@article{Born1938ASF,
  author        = {Max Born},
  title         = {A suggestion for unifying quantum theory and relativity},
  journal       = {Proceedings of the Royal Society of London. Series A. Mathematical and Physical Sciences},
  volume        = {165},
  pages         = {291 - 303},
  year          = {1938}
}

@article{Bousso:2002ju,
  author        = "Bousso, Raphael",
  title         = "{The Holographic principle}",
  journal       = "Rev. Mod. Phys.",
  volume        = "74",
  pages         = "825--874",
  year          = "2002",
  eprint        = "hep-th/0203101",
  archivePrefix = "arXiv",
  reportNumber  = "NSF-ITP-02-17",
  doi           = "10.1103/RevModPhys.74.825"
}

@article{brandenberger1989superstrings,
  author        = {Brandenberger, Robert and Vafa, Cumrun},
  title         = {Superstrings in the early universe},
  journal       = {Nuclear Physics B},
  volume        = {316},
  number        = {2},
  pages         = {391--410},
  year          = {1989},
  publisher     = {Elsevier}
}

@article{Brandenberger:2008nx,
  author        = "Brandenberger, Robert H.",
  title         = "{String Gas Cosmology}",
  journal       = "{arXiv preprint}",
  year          = "2008",
  month         = aug,
  eprint        = "0808.0746",
  archivePrefix = "arXiv",
  primaryClass  = "hep-th"
}

@article{Carballo-Rubio:2022aed,
  author        = "Carballo-Rubio, Ra\'ul and Cardoso, Vitor and Younsi, Ziri",
  title         = "{Toward very large baseline interferometry observations of black hole structure}",
  journal       = "Phys. Rev. D",
  volume        = "106",
  number        = "8",
  pages         = "084038",
  year          = "2022",
  eprint        = "2208.00704",
  archivePrefix = "arXiv",
  primaryClass  = "gr-qc",
  doi           = "10.1103/PhysRevD.106.084038"
}

@article{Carballo-Rubio:2023fjj,
  author        = "Carballo-Rubio, Ra\'ul and Di Filippo, Francesco and Liberati, Stefano and Visser, Matt",
  title         = "{Constraints on thermalizing surfaces from infrared observations of supermassive black holes}",
  journal       = "JCAP",
  volume        = "11",
  pages         = "041",
  year          = "2023",
  eprint        = "2306.17480",
  archivePrefix = "arXiv",
  primaryClass  = "astro-ph.HE",
  reportNumber  = "YITP-22-84",
  doi           = "10.1088/1475-7516/2023/11/041"
}

@article{Carballo-Rubio:2023mvr,
  author        = "Carballo-Rubio, Ra\'ul and Di Filippo, Francesco and Liberati, Stefano and Visser, Matt",
  title         = "{Singularity-free gravitational collapse: From regular black holes to horizonless objects}",
  year          = "2023",
  month         = jan,
  eprint        = "2302.00028",
  archivePrefix = "arXiv",
  primaryClass  = "gr-qc"
}

@article{Carballo-Rubio:2025fnc,
  author        = "Carballo-Rubio, Ra\'ul and others",
  title         = "{Towards a Non-singular Paradigm of Black Hole Physics}",
  year          = "2025",
  month         = jan,
  eprint        = "2501.05505",
  archivePrefix = "arXiv",
  primaryClass  = "gr-qc"
}

@article{Cardoso:2016rao,
  author        = "Cardoso, Vitor and Franzin, Edgardo and Pani, Paolo",
  title         = "{Is the gravitational-wave ringdown a probe of the event horizon?}",
  journal       = "Phys. Rev. Lett.",
  volume        = "116",
  number        = "17",
  pages         = "171101",
  year          = "2016",
  eprint        = "1602.07309",
  archivePrefix = "arXiv",
  primaryClass  = "gr-qc",
  doi           = "10.1103/PhysRevLett.116.171101",
  note          = "[Erratum: Phys.Rev.Lett. 117, 089902 (2016)]"
}

@article{Cardoso:2017cqb,
  author        = "Cardoso, Vitor and Pani, Paolo",
  title         = "{Tests for the existence of black holes through gravitational wave echoes}",
  journal       = "Nature Astron.",
  volume        = "1",
  number        = "9",
  pages         = "586--591",
  year          = "2017",
  eprint        = "1709.01525",
  archivePrefix = "arXiv",
  primaryClass  = "gr-qc",
  doi           = "10.1038/s41550-017-0225-y"
}

@article{CardosoPani2019,
  author        = {Cardoso, Vitor and Pani, Paolo},
  title         = {Testing the Nature of Dark Compact Objects},
  journal       = {Living Reviews in Relativity},
  volume        = {22},
  number        = {1},
  pages         = {4},
  year          = {2019},
  eprint        = {1904.05363},
  archivePrefix = {arXiv},
  primaryClass  = {gr-qc},
  doi           = {10.1007/s41114-019-0020-4}
}

@article{Carr:2014mya,
  author        = "Carr, B. J.",
  editor        = "Nicolini, Piero and Kaminski, Matthias and Mureika, Jonas and Bleicher, Marcus",
  title         = "{The Black Hole Uncertainty Principle Correspondence}",
  journal       = "Springer Proc. Phys.",
  volume        = "170",
  pages         = "159--167",
  year          = "2016",
  eprint        = "1402.1427",
  archivePrefix = "arXiv",
  primaryClass  = "gr-qc",
  doi           = "10.1007/978-3-319-20046-0_19"
}

@article{Carr:2015nqa,
  author        = "Carr, Bernard J. and Mureika, Jonas and Nicolini, Piero",
  title         = "{Sub-Planckian black holes and the Generalized Uncertainty Principle}",
  journal       = "JHEP",
  volume        = "07",
  pages         = "052",
  year          = "2015",
  eprint        = "1504.07637",
  archivePrefix = "arXiv",
  primaryClass  = "gr-qc",
  doi           = "10.1007/JHEP07(2015)052"
}

@article{Carr:2020gox,
  author        = "Carr, Bernard and Kohri, Kazunori and Sendouda, Yuuiti and Yokoyama, Jun'ichi",
  title         = "Constraints on primordial black holes",
  journal       = "Rept. Prog. Phys.",
  volume        = "84",
  number        = "11",
  pages         = "116902",
  year          = "2021",
  eprint        = "2002.12778",
  archivePrefix = "arXiv",
  primaryClass  = "astro-ph.CO",
  doi           = "10.1088/1361-6633/ac1e31"
}

@article{Chamseddine:2019pux,
  author        = "Chamseddine, Ali H. and Mukhanov, Viatcheslav and Russ, Tobias B.",
  title         = "{Black Hole Remnants}",
  journal       = "JHEP",
  volume        = "10",
  pages         = "104",
  year          = "2019",
  eprint        = "1908.03498",
  archivePrefix = "arXiv",
  primaryClass  = "hep-th",
  doi           = "10.1007/JHEP10(2019)104"
}

@article{ChamseddineMukhanov2013,
  author        = {Chamseddine, Ali H. and Mukhanov, Viatcheslav},
  title         = {Mimetic Dark Matter},
  journal       = {Journal of High Energy Physics},
  volume        = {2013},
  number        = {11},
  pages         = {135},
  year          = {2013},
  eprint        = {1308.5410},
  archivePrefix = {arXiv},
  primaryClass  = {astro-ph.CO},
  doi           = {10.1007/JHEP11(2013)135}
}

@article{ChamseddineMukhanov2017,
  author        = {Chamseddine, Ali H. and Mukhanov, Viatcheslav},
  title         = {Resolving Cosmological Singularities},
  journal       = {Journal of Cosmology and Astroparticle Physics},
  volume        = {2017},
  number        = {03},
  pages         = {009},
  year          = {2017},
  eprint        = {1612.05860},
  archivePrefix = {arXiv},
  primaryClass  = {gr-qc},
  doi           = {10.1088/1475-7516/2017/03/009}
}

@article{ChamseddineMukhanov2017LC,
  author        = {Chamseddine, Ali H. and Mukhanov, Viatcheslav},
  title         = {Nonsingular Black Hole},
  journal       = {European Physical Journal C},
  volume        = {77},
  number        = {3},
  pages         = {183},
  year          = {2017},
  eprint        = {1612.05861},
  archivePrefix = {arXiv},
  primaryClass  = {gr-qc},
  doi           = {10.1140/epjc/s10052-017-4779-5}
}

@article{Chen:2002tu,
  author        = "Chen, Pisin and Adler, Ronald J.",
  editor        = "Cline, D. B.",
  title         = "{Black hole remnants and dark matter}",
  journal       = "Nucl. Phys. B Proc. Suppl.",
  volume        = "124",
  pages         = "103--106",
  year          = "2003",
  eprint        = "gr-qc/0205106",
  archivePrefix = "arXiv",
  reportNumber  = "SLAC-PUB-9331",
  doi           = "10.1016/S0920-5632(03)02088-7"
}

@article{Chouha-PR-cosm:2026,
  author        = "Robert Brandenberger and Chouha, Paul-Robert",
  title         = "{Metaparticles and Cosmology (In Preparation)}"
}

@article{Chouha-PR-cosm:2026-b,
  author        = "Chouha, Paul-Robert",
  title         = "{Non-Singular Black Hole Remnant with a Non-Geomteric Stabel Core. (In Preparation)}"
}

@article{Chouha-PR-cosm:2026-c,
  author        = "Chouha, Paul-Robert",
  title         = "{Extended Generalized Uncertainty Principle from Metaparticles and Metastrings.(In Preparation)}"
}

@article{Chouha-PR:2025,
  author        = "Chouha, Paul-Robert",
  title         = "{An Effective Metric for the Metaparticle-Non-singular Black Hole (In Preparation)}"
}

@article{ChristodoulouRovelli2015,
  author        = {Christodoulou, Marios and Rovelli, Carlo},
  title         = {How Big Is a Black Hole?},
  journal       = {Physical Review D},
  volume        = {91},
  number        = {6},
  pages         = {064046},
  year          = {2015},
  eprint        = {1411.2854},
  archivePrefix = {arXiv},
  primaryClass  = {gr-qc},
  doi           = {10.1103/PhysRevD.91.064046}
}

@article{Corless1996LambertW,
  author        = {Corless, R. M. and Gonnet, G. H. and Hare, D. E. G. and Jeffrey, D. J. and Knuth, D. E.},
  title         = {On the Lambert W Function},
  journal       = {Advances in Computational Mathematics},
  volume        = {5},
  pages         = {329--359},
  year          = {1996},
  doi           = {10.1007/BF02124750}
}

@article{DebyeHuckel1923,
  author        = {Debye, P. and H{\"u}ckel, E.},
  title         = {Zur Theorie der Elektrolyte},
  journal       = {Physikalische Zeitschrift},
  volume        = {24},
  pages         = {185--206},
  year          = {1923}
}

@article{Eichhorn:2022bbn,
  author        = "Eichhorn, Astrid and Held, Aaron",
  title         = "{Quantum gravity lights up spinning black holes}",
  journal       = "JCAP",
  volume        = "01",
  pages         = "032",
  year          = "2023",
  eprint        = "2206.11152",
  archivePrefix = "arXiv",
  primaryClass  = "gr-qc",
  doi           = "10.1088/1475-7516/2023/01/032"
}

@article{EventHorizonTelescope:2022xqj,
  author        = "Akiyama, Kazunori and others",
  collaboration = "Event Horizon Telescope",
  title         = "{First Sagittarius A* Event Horizon Telescope Results. VI. Testing the Black Hole Metric}",
  journal       = "Astrophys. J. Lett.",
  volume        = "930",
  number        = "2",
  pages         = "L17",
  year          = "2022",
  eprint        = "2311.09484",
  archivePrefix = "arXiv",
  primaryClass  = "astro-ph.HE",
  reportNumber  = "FERMILAB-PUB-22-422-PPD",
  doi           = "10.3847/2041-8213/ac6756"
}

@book{FabbriNavarro,
  author        = {Fabbri, Alessandro and Navarro-Salas, Jos{\'e}},
  title         = {Modeling Black Hole Evaporation},
  year          = {2005},
  publisher     = {Imperial College Press}
}

@article{Freidel:2015pka,
  author        = "Freidel, Laurent and Leigh, Robert G. and Minic, Djordje",
  title         = "{Metastring Theory and Modular Space-time}",
  journal       = "JHEP",
  volume        = "06",
  pages         = "006",
  year          = "2015",
  eprint        = "1502.08005",
  archivePrefix = "arXiv",
  primaryClass  = "hep-th",
  doi           = "10.1007/JHEP06(2015)006"
}

@article{Freidel:2015uug,
  author        = "Freidel, Laurent and Leigh, Robert G. and Minic, Djordje",
  title         = "{Modular spacetime}",
  journal       = "Int. J. Mod. Phys. D",
  volume        = "24",
  number        = "12",
  pages         = "1544028",
  year          = "2015",
  doi           = "10.1142/S0218271815440289"
}

@article{Freidel:2016pls,
  author        = "Freidel, Laurent and Leigh, Robert G. and Minic, Djordje",
  title         = "{Quantum Spaces are Modular}",
  journal       = "Phys. Rev. D",
  volume        = "94",
  number        = "10",
  pages         = "104052",
  year          = "2016",
  eprint        = "1606.01829",
  archivePrefix = "arXiv",
  primaryClass  = "hep-th",
  doi           = "10.1103/PhysRevD.94.104052"
}

@article{Freidel:2017xsi,
  author        = "Freidel, Laurent and Leigh, Robert G. and Minic, Djordje",
  editor        = "Burd\'\i{}k, \v{C}estm\'\i{}r and Navr\'atil, Ond\v{r}ej and Po\v{s}ta, Severin",
  title         = "{Modular Spacetime and Metastring Theory}",
  journal       = "J. Phys. Conf. Ser.",
  volume        = "804",
  number        = "1",
  pages         = "012032",
  year          = "2017",
  doi           = "10.1088/1742-6596/804/1/012032"
}

@article{Freidel:2018apz,
  author        = "Freidel, Laurent and Kowalski-Glikman, Jerzy and Leigh, Robert G. and Minic, Djordje",
  title         = "{Theory of metaparticles}",
  journal       = "Phys. Rev. D",
  volume        = "99",
  number        = "6",
  pages         = "066011",
  year          = "2019",
  eprint        = "1812.10821",
  archivePrefix = "arXiv",
  primaryClass  = "hep-th",
  doi           = "10.1103/PhysRevD.99.066011"
}

@article{Freidel:2021wpl,
  author        = "Freidel, Laurent and Kowalski-Glikman, Jerzy and Leigh, Robert G. and Minic, Djordje",
  title         = "{Quantum gravity phenomenology in the infrared}",
  journal       = "Int. J. Mod. Phys. D",
  volume        = "30",
  number        = "14",
  pages         = "2141002",
  year          = "2021",
  eprint        = "2104.00802",
  archivePrefix = "arXiv",
  primaryClass  = "gr-qc",
  doi           = "10.1142/S0218271821410029"
}

@article{FreidelLeighMinic2015,
  author        = {Freidel, Laurent and Leigh, Robert G. and Minic, Djordje},
  title         = {Metastring Theory and Modular Spacetime},
  journal       = {Journal of High Energy Physics},
  volume        = {2015},
  number        = {06},
  pages         = {006},
  year          = {2015},
  eprint        = {1502.08005},
  archivePrefix = {arXiv},
  primaryClass  = {hep-th},
  doi           = {10.1007/JHEP06(2015)006}
}

@article{Gibbons:1975kk,
  author        = "Gibbons, G. W.",
  title         = "{Vacuum Polarization and the Spontaneous Loss of Charge by Black Holes}",
  journal       = "Commun. Math. Phys.",
  volume        = "44",
  pages         = "245--264",
  year          = "1975",
  doi           = "10.1007/BF01609829"
}

@article{gliner1969regular,
  author        = {E. Gliner},
  title         = {On the possibility of creating a regular black hole},
  journal       = {Zh. Eksp. Teor. Fiz.},
  volume        = {56},
  pages         = {1781},
  year          = {1969},
  translation   = {Sov. Phys. JETP 22, 1298 (1966)}
}

@article{HaggardRovelli2015,
  author        = {Haggard, Hal M. and Rovelli, Carlo},
  title         = {Quantum-Gravity Effects Outside the Horizon Spark Black to White Hole Tunneling},
  journal       = {Physical Review D},
  volume        = {92},
  number        = {10},
  pages         = {104020},
  year          = {2015},
  eprint        = {1407.0989},
  archivePrefix = {arXiv},
  primaryClass  = {gr-qc},
  doi           = {10.1103/PhysRevD.92.104020}
}

@article{hawking1969singularity,
  author        = {Stephen W. Hawking and Roger Penrose},
  title         = {The Singularities of Gravitational Collapse and Cosmology},
  journal       = {Proceedings of the Royal Society of London. Series A, Mathematical and Physical Sciences},
  volume        = {314},
  number        = {1519},
  pages         = {529--548},
  year          = {1969},
  doi           = {10.1098/rspa.1969.0170}
}

@article{Hawking1974a,
  author        = {S. W. Hawking},
  title         = {Black Hole explosions?},
  journal       = {Nature},
  volume        = {248},
  pages         = {30--31},
  year          = {1974},
  doi           = {10.1038/248030a0},
  url           = {https://doi.org/10.1038/248030a0}
}

@article{Hawking1975,
  author        = {S. W. Hawking},
  title         = {Particle creation by black holes},
  journal       = {Communications in Mathematical Physics},
  volume        = {43},
  number        = {3},
  pages         = {199--220},
  year          = {1975},
  doi           = {10.1007/BF02345020},
  url           = {https://doi.org/10.1007/BF02345020}
}

@article{Hiscock1981,
  author        = {Hiscock, William A.},
  title         = {Models of evaporating black holes. I. Semi-classical solutions},
  journal       = {Phys. Rev. D},
  volume        = {23},
  pages         = {2813--2822},
  year          = {1981}
}

@article{Hossenfelder:2005ed,
  author        = "Hossenfelder, S.",
  title         = "{Self-consistency in theories with a minimal length}",
  journal       = "Class. Quant. Grav.",
  volume        = "23",
  pages         = "1815--1821",
  year          = "2006",
  eprint        = "hep-th/0510245",
  archivePrefix = "arXiv",
  doi           = "10.1088/0264-9381/23/5/N01"
}

@article{Hull2005,
  author        = {Hull, C. M.},
  title         = {A Geometry for Non-Geometric String Backgrounds},
  journal       = {Journal of High Energy Physics},
  volume        = {2005},
  number        = {10},
  pages         = {065},
  year          = {2005},
  eprint        = {hep-th/0406102},
  archivePrefix = {arXiv},
  doi           = {10.1088/1126-6708/2005/10/065}
}

@article{Hull:2009mi,
  author        = "Hull, Chris and Zwiebach, Barton",
  title         = "{Double Field Theory}",
  journal       = "JHEP",
  volume        = "09",
  pages         = "099",
  year          = "2009",
  eprint        = "0904.4664",
  archivePrefix = "arXiv",
  primaryClass  = "hep-th",
  reportNumber  = "IMPERIAL-TP-2009-CH-02, MIT-CTP-4031",
  doi           = "10.1088/1126-6708/2009/09/099"
}

@article{Jusufi:2024dtr,
  author        = "Jusufi, Kimet and Nicolini, Piero",
  title         = "{Geodesic completeness from string T-duality}",
  year          = "2024",
  month         = oct,
  eprint        = "2410.19613",
  archivePrefix = "arXiv",
  primaryClass  = "hep-th"
}

@article{Lemos:2008cv,
  author        = "Lemos, Jose P. S. and Zaslavskii, Oleg B.",
  title         = "{Black hole mimickers: Regular versus singular behavior}",
  journal       = "Phys. Rev. D",
  volume        = "78",
  pages         = "024040",
  year          = "2008",
  eprint        = "0806.0845",
  archivePrefix = "arXiv",
  primaryClass  = "gr-qc",
  doi           = "10.1103/PhysRevD.78.024040"
}

@book{Lifshitz,
  author        = {Lifshitz E M, Pitaevskii L P and Berestetskii},
  title         = {Course of Theoretical Physics vol 4 Quantum Electrodynamics Landau-Lifshitz},
  year          = {1982},
  publisher     = {Oxford: Reed Educational and Professional Publishing.}
}

@article{Maldacena:2001kr,
  author        = "Maldacena, Juan Martin",
  title         = "{Eternal black holes in anti-de Sitter}",
  journal       = "JHEP",
  volume        = "04",
  pages         = "021",
  year          = "2003",
  eprint        = "hep-th/0106112",
  archivePrefix = "arXiv",
  reportNumber  = "NSF-ITP-01-59",
  doi           = "10.1088/1126-6708/2003/04/021"
}

@article{Mathur2005,
  author        = {Mathur, Samir D.},
  title         = {The Fuzzball Proposal for Black Holes},
  journal       = {Fortschritte der Physik},
  volume        = {53},
  pages         = {793--827},
  year          = {2005},
  eprint        = {hep-th/0502050},
  archivePrefix = {arXiv},
  doi           = {10.1002/prop.200410203}
}

@article{Mathur2009,
  author        = {Mathur, Samir D.},
  title         = {The Information Paradox: A Pedagogical Introduction},
  journal       = {Classical and Quantum Gravity},
  volume        = {26},
  pages         = {224001},
  year          = {2009},
  eprint        = {0909.1038},
  archivePrefix = {arXiv},
  primaryClass  = {hep-th},
  doi           = {10.1088/0264-9381/26/22/224001}
}

@article{Mazur:2015kia,
  author        = "Mazur, Pawel O. and Mottola, Emil",
  title         = "{Surface tension and negative pressure interior of a non-singular \textquoteleft{}black hole\textquoteright{}}",
  journal       = "Class. Quant. Grav.",
  volume        = "32",
  number        = "21",
  pages         = "215024",
  year          = "2015",
  eprint        = "1501.03806",
  archivePrefix = "arXiv",
  primaryClass  = "gr-qc",
  reportNumber  = "LA-UR-15-20030",
  doi           = "10.1088/0264-9381/32/21/215024"
}

@article{MazurMottola2004,
  author        = {Mazur, Pawel O. and Mottola, Emil},
  title         = {Gravitational Condensate Stars: An Alternative to Black Holes},
  journal       = {Proceedings of the National Academy of Sciences},
  volume        = {101},
  pages         = {9545--9550},
  year          = {2004},
  eprint        = {gr-qc/0407075},
  archivePrefix = {arXiv},
  doi           = {10.1073/pnas.0402717101}
}

@article{MukhanovBrandenberger1992,
  author        = {Mukhanov, Viatcheslav and Brandenberger, Robert},
  title         = {A Nonsingular Universe},
  journal       = {Physical Review Letters},
  volume        = {68},
  pages         = {1969--1972},
  year          = {1992},
  doi           = {10.1103/PhysRevLett.68.1969}
}

@article{Murk:2023rwl,
  author        = "Murk, Sebastian and Soranidis, Ioannis",
  title         = "{Regular black holes and the first law of black hole mechanics}",
  journal       = "Phys. Rev. D",
  volume        = "108",
  number        = "4",
  pages         = "044002",
  year          = "2023",
  eprint        = "2304.05421",
  archivePrefix = "arXiv",
  primaryClass  = "gr-qc",
  doi           = "10.1103/PhysRevD.108.044002"
}

@article{Nicolini2019,
  author        = {Nicolini, Piero and Wondrak, Michael and Gaete, Patricio},
  title         = {Quantum Corrected Black Holes from String T-Duality},
  journal       = {arXiv preprint arXiv:1902.11242},
  year          = {2019},
  url           = {https://arxiv.org/abs/1902.11242}
}

@article{Nicolini:2005vd,
  author        = "Nicolini, Piero and Smailagic, Anais and Spallucci, Euro",
  title         = "{Noncommutative geometry inspired Schwarzschild black hole}",
  journal       = "Phys. Lett. B",
  volume        = "632",
  pages         = "547--551",
  year          = "2006",
  eprint        = "gr-qc/0510112",
  archivePrefix = "arXiv",
  doi           = "10.1016/j.physletb.2005.11.004"
}

@article{Nozari:2006bi,
  author        = "Nozari, Kourosh and Fazlpour, Behnaz",
  title         = "{Thermodynamics of an evaporating Schwarzschild black hole in noncommutative space}",
  journal       = "Mod. Phys. Lett. A",
  volume        = "22",
  pages         = "2917--2930",
  year          = "2007",
  eprint        = "hep-th/0605109",
  archivePrefix = "arXiv",
  doi           = "10.1142/S0217732307023602"
}

@article{OlmoRubieraGarcia2015,
  author        = {Olmo, Gonzalo J. and Rubiera-Garcia, Diego},
  title         = {Nonsingular Black Holes in $f(R)$ Theories},
  journal       = {Physical Review D},
  volume        = {92},
  number        = {4},
  pages         = {044047},
  year          = {2015},
  eprint        = {1507.07777},
  archivePrefix = {arXiv},
  primaryClass  = {gr-qc},
  doi           = {10.1103/PhysRevD.92.044047}
}

@article{Padmanabhan:1996ap,
  author        = "Padmanabhan, T.",
  title         = "{Duality and zero point length of space-time}",
  journal       = "Phys. Rev. Lett.",
  volume        = "78",
  pages         = "1854--1857",
  year          = "1997",
  eprint        = "hep-th/9608182",
  archivePrefix = "arXiv",
  reportNumber  = "IUCAA-35-96",
  doi           = "10.1103/PhysRevLett.78.1854"
}

@article{Page1976,
  author        = {Page, Don N.},
  title         = {Particle emission rates from a black hole: Massless particles from an uncharged, nonrotating hole},
  journal       = {Phys. Rev. D},
  volume        = {13},
  pages         = {198--206},
  year          = {1976}
}

@article{Page:1976df,
  author        = "Page, Don N.",
  title         = "{Particle Emission Rates from a Black Hole: Massless Particles from an Uncharged, Nonrotating Hole}",
  journal       = "Phys. Rev. D",
  volume        = "13",
  pages         = "198--206",
  year          = "1976",
  doi           = "10.1103/PhysRevD.13.198"
}

@article{Parentani1994,
  author        = {Parentani, Roberto and Piran, Tsvi},
  title         = {The internal geometry of an evaporating black hole},
  journal       = {Phys. Rev. Lett.},
  volume        = {73},
  pages         = {2805--2808},
  year          = {1994}
}

@article{RovelliVidotto2014,
  author        = {Rovelli, Carlo and Vidotto, Francesca},
  title         = {Planck Stars},
  journal       = {International Journal of Modern Physics D},
  volume        = {23},
  number        = {12},
  pages         = {1442026},
  year          = {2014},
  eprint        = {1401.6562},
  archivePrefix = {arXiv},
  primaryClass  = {gr-qc},
  doi           = {10.1142/S0218271814420267}
}

@article{Scardigli:1999jh,
  author        = "Scardigli, Fabio",
  title         = "{Generalized uncertainty principle in quantum gravity from micro - black hole Gedanken experiment}",
  journal       = "Phys. Lett. B",
  volume        = "452",
  pages         = "39--44",
  year          = "1999",
  eprint        = "hep-th/9904025",
  archivePrefix = "arXiv",
  doi           = "10.1016/S0370-2693(99)00167-7"
}

@article{Schwinger:1951nm,
  author        = "Schwinger, Julian S.",
  editor        = "Milton, K. A.",
  title         = "{On gauge invariance and vacuum polarization}",
  journal       = "Phys. Rev.",
  volume        = "82",
  pages         = "664--679",
  year          = "1951",
  doi           = "10.1103/PhysRev.82.664"
}

@article{Smailagic:2003hm,
  author        = "Smailagic, A. and Spallucci, E. and Padmanabhan, T.",
  title         = "{String theory T duality and the zero point length of space-time}",
  year          = "2003",
  month         = aug,
  eprint        = "hep-th/0308122",
  archivePrefix = "arXiv"
}

@article{Tseytlin:1990nb,
  author        = "Tseytlin, Arkady A.",
  title         = "{Duality Symmetric Formulation of String
    World Sheet Dynamics}",
  journal       = "Phys. Lett. B",
  volume        = "242",
  pages         = "163--174",
  year          = "1990",
  reportNumber  = "KCL-TP-1990-2",
  doi           = "10.1016/0370-2693(90)91454-J"
}

@article{VanRaamsdonk:2010pw,
  author        = "Van Raamsdonk, Mark",
  title         = "{Building up spacetime with quantum entanglement}",
  journal       = "Gen. Rel. Grav.",
  volume        = "42",
  pages         = "2323--2329",
  year          = "2010",
  eprint        = "1005.3035",
  archivePrefix = "arXiv",
  primaryClass  = "hep-th",
  doi           = "10.1142/S0218271810018529"
}

@article{Volovik:2021upi,
  author        = "Volovik, G. E.",
  title         = {{Effect of the inner horizon on the black hole thermodynamics: Reissner{\textendash}Nordstr{\"o}m black hole and Kerr black hole}},
  journal       = "Mod. Phys. Lett. A",
  volume        = "36",
  number        = "24",
  pages         = "2150177",
  year          = "2021",
  eprint        = "2107.11193",
  archivePrefix = "arXiv",
  primaryClass  = "gr-qc",
  doi           = "10.1142/S0217732321501777"
}

@article{Rovelli1996Immirzi,
  author = {Rovelli, Carlo and Thiemann, Thomas},
  title = {The Immirzi parameter in quantum general relativity},
  journal = {Physical Review D},
  volume = {57},
  number = {2},
  pages = {1009--1014},
  year = {1998},
  doi = {10.1103/PhysRevD.57.1009},
  eprint = {gr-qc/9705059},
  archivePrefix = {arXiv}
}

@article{AshtekarBaezKrasnov2000,
  author = {Ashtekar, Abhay and Baez, John and Krasnov, Kirill},
  title = {Quantum geometry of isolated horizons and black hole entropy},
  journal = {Advances in Theoretical and Mathematical Physics},
  volume = {4},
  pages = {1--94},
  year = {2000},
  eprint = {gr-qc/0005126},
  archivePrefix = {arXiv}
}

@article{Spallucci:2008ez,
    author = "Spallucci, Euro and Smailagic, Anais and Nicolini, Piero",
    title = "{Non-commutative geometry inspired higher-dimensional charged black holes}",
    eprint = "0801.3519",
    archivePrefix = "arXiv",
    primaryClass = "hep-th",
    doi = "10.1016/j.physletb.2008.11.030",
    journal = "Phys. Lett. B",
    volume = "670",
    pages = "449--454",
    year = "2009"
}

@article{Dvali:2020wqi,
  author = {Dvali, Gia and Gomez, Cesar},
  title = {Memory burden and black hole information},
  eprint = {2003.03679},
  archivePrefix = {arXiv},
  primaryClass = {hep-th},
  journal = {Eur. Phys. J. C},
  volume = {81},
  number = {4},
  pages = {349},
  year = {2021}
}

@article{Cardoso:2019rvt,
  author = {Cardoso, Vitor and Pani, Paolo},
  title = {Testing the nature of dark compact objects: a status report},
  eprint = {1904.05363},
  archivePrefix = {arXiv},
  primaryClass = {gr-qc},
  journal = {Living Rev. Rel.},
  volume = {22},
  number = {1},
  pages = {4},
  year = {2019}
}

@article{Gralla:2025gzl,
    author = "Gralla, Samuel E.",
    title = "{Can black holes evaporate past extremality?}",
    eprint = "2510.18839",
    archivePrefix = "arXiv",
    primaryClass = "gr-qc",
    month = "10",
    year = "2025"
}

@article{Friedmann:1922,
  author  = {Friedmann, Alexander},
  title   = {On the Curvature of Space},
  journal = {Zeitschrift f{\"u}r Physik},
  volume  = {10},
  pages   = {377--386},
  year    = {1922}
}

@article{Friedmann:1924,
  author  = {Friedmann, Alexander},
  title   = {On the Possibility of a World with Constant Negative Curvature of Space},
  journal = {Zeitschrift f{\"u}r Physik},
  volume  = {21},
  pages   = {326--332},
  year    = {1924}
}

@article{Lemaitre:1927,
  author  = {Lema{\^i}tre, Georges},
  title   = {Un Univers homog{\`e}ne de masse constante et de rayon croissant, rendant compte de la vitesse radiale des n{\'e}buleuses extra-galactiques},
  journal = {Annales de la Soci{\'e}t{\'e} Scientifique de Bruxelles},
  volume  = {47},
  pages   = {49--59},
  year    = {1927}
}

@article{Robertson:1935,
  author  = {Robertson, Howard P.},
  title   = {Kinematics and World-Structure},
  journal = {Astrophysical Journal},
  volume  = {82},
  pages   = {284--301},
  year    = {1935}
}

@article{Walker:1937,
  author  = {Walker, Arthur G.},
  title   = {On Milne's Theory of World-Structure},
  journal = {Proceedings of the London Mathematical Society},
  volume  = {42},
  pages   = {90--127},
  year    = {1937}
}

@article{GrossMende1988,
  author        = {Gross, David J. and Mende, Paul F.},
  title         = {String Theory Beyond the Planck Scale},
  journal       = {Nuclear Physics B},
  volume        = {303},
  pages         = {407--454},
  year          = {1988},
  doi           = {10.1016/0550-3213(88)90390-2}
}

@article{AmatiCiafaloniVeneziano1989,
  author        = {Amati, Daniele and Ciafaloni, Marcello and Veneziano, Gabriele},
  title         = {Can Space-Time Be Probed Below the String Size?},
  journal       = {Physics Letters B},
  volume        = {216},
  pages         = {41--47},
  year          = {1989},
  doi           = {10.1016/0370-2693(89)91366-X}
}

@article{KempfManganoMann1995,
  author        = {Kempf, Achim and Mangano, Gianpiero and Mann, Robert B.},
  title         = {Hilbert Space Representation of the Minimal Length Uncertainty Relation},
  journal       = {Physical Review D},
  volume        = {52},
  pages         = {1108--1118},
  year          = {1995},
  doi           = {10.1103/PhysRevD.52.1108}
}

@article{RovelliSmolin1995,
  author        = {Rovelli, Carlo and Smolin, Lee},
  title         = {Discreteness of Area and Volume in Quantum Gravity},
  journal       = {Nuclear Physics B},
  volume        = {442},
  pages         = {593--622},
  year          = {1995},
  doi           = {10.1016/0550-3213(95)00150-Q}
}

@article{DoplicherFredenhagenRoberts1995,
  author        = {Doplicher, Sergio and Fredenhagen, Klaus and Roberts, John E.},
  title         = {The Quantum Structure of Space-Time at the Planck Scale and Quantum Fields},
  journal       = {Communications in Mathematical Physics},
  volume        = {172},
  pages         = {187--220},
  year          = {1995},
  doi           = {10.1007/BF02104515}
}

@article{Mead1964,
  author        = {Mead, C. Alden},
  title         = {Possible Connection Between Gravitation and Fundamental Length},
  journal       = {Physical Review},
  volume        = {135},
  pages         = {B849--B862},
  year          = {1964},
  doi           = {10.1103/PhysRev.135.B849}
}

@article{Garay1995,
  author        = {Garay, Luis J.},
  title         = {Quantum Gravity and Minimum Length},
  journal       = {International Journal of Modern Physics A},
  volume        = {10},
  pages         = {145--165},
  year          = {1995},
  doi           = {10.1142/S0217751X95000085}
}

@article{Hossenfelder2013,
  author        = {Hossenfelder, Sabine},
  title         = {Minimal Length Scale Scenarios for Quantum Gravity},
  journal       = {Living Reviews in Relativity},
  volume        = {16},
  pages         = {2},
  year          = {2013},
  doi           = {10.12942/lrr-2013-2}
}

@book{Polchinski:1998rq,
  author       = {Polchinski, Joseph},
  title        = {String Theory. Vol. 1: An Introduction to the Bosonic String},
  publisher    = {Cambridge University Press},
  year         = {1998},
  isbn         = {978-0521672276}
}

@book{Green:1987sp,
  author       = {Green, Michael B. and Schwarz, John H. and Witten, Edward},
  title        = {Superstring Theory. Vol. 1: Introduction},
  publisher    = {Cambridge University Press},
  year         = {1987},
  isbn         = {978-0521352529}
}

@article{KaulMajumdar2000,
  author        = {Kaul, R. K. and Majumdar, P.},
  title         = {Logarithmic Correction to the Bekenstein-Hawking Entropy},
  journal       = {Phys. Rev. Lett.},
  volume        = {84},
  pages         = {5255},
  year          = {2000},
  doi           = {10.1103/PhysRevLett.84.5255},
  eprint        = {gr-qc/0002040},
  archivePrefix = {arXiv}
}

@article{Sen2013Log,
  author        = {Sen, Ashoke},
  title         = {Logarithmic Corrections to Schwarzschild and Other Non-extremal Black Hole Entropy in Different Dimensions},
  journal       = {JHEP},
  volume        = {04},
  pages         = {156},
  year          = {2013},
  doi           = {10.1007/JHEP04(2013)156},
  eprint        = {1205.0971},
  archivePrefix = {arXiv},
  primaryClass  = {hep-th}
}

@article{Solodukhin2011,
  author        = {Solodukhin, Sergey N.},
  title         = {Entanglement Entropy of Black Holes},
  journal       = {Living Rev. Rel.},
  volume        = {14},
  pages         = {8},
  year          = {2011},
  doi           = {10.12942/lrr-2011-8},
  eprint        = {1104.3712},
  archivePrefix = {arXiv}
}

@article{Fursaev1995,
  author        = {Fursaev, Dmitri V.},
  title         = {Temperature and entropy of a quantum black hole and conformal anomaly},
  journal       = {Phys. Rev. D},
  volume        = {51},
  pages         = {5352},
  year          = {1995},
  doi           = {10.1103/PhysRevD.51.5352},
  eprint        = {hep-th/9412161},
  archivePrefix = {arXiv}
}

@article{https://doi.org/10.1111/j.1749-6632.1973.tb41445.x,
author = {Geroch, Robert},
title = {ENERGY EXTRACTION},
journal = {Annals of the New York Academy of Sciences},
volume = {224},
number = {1},
pages = {108-117},
doi = {https://doi.org/10.1111/j.1749-6632.1973.tb41445.x},
url = {https://nyaspubs.onlinelibrary.wiley.com/doi/abs/10.1111/j.1749-6632.1973.tb41445.x},
eprint = {https://nyaspubs.onlinelibrary.wiley.com/doi/pdf/10.1111/j.1749-6632.1973.tb41445.x},
year = {1973}
}

@article{landauer1961irreversibility,
  title={Irreversibility and heat generation in the computing process},
  author={Landauer, Rolf},
  journal={IBM Journal of Research and Development},
  volume={5},
  number={3},
  pages={183--191},
  year={1961},
  publisher={IBM}
}
\end{document}